\documentclass[11pt,preprint]{aastex}

\def\persqcm{$\rm cm^{-2}$}

\def\htoo{$\rm H_2$}
\def\h2{$\rm H_2$}
\def\error{$\pm$}
\def\e#1{$\times 10^{#1}$}
\def\tenup#1{10$^{#1}$}
\def\asec{\arcsec}

\def\deg{\arcdeg}

\def\etal{et al.~}
\def\kms{km~s$^{-1}$}

\def\solmass{$\rm M_{\sun}$}
\def\solum{$\rm L_{\sun}$}

\newcommand{\htooco}{$\rm H_2/CO$}
\newcommand{\convunits}{$\rm cm^{-2}\,(K\,km\,s^{-1})^{-1}$}

\newcommand{\jybks}{$\rm Jy\,b^{-1}\,km\,s^{-1}$}
\newcommand{\jykms}{$\rm Jy\,km\,s^{-1}$}
\newcommand{\mjb}{$\rm mJy\:beam^{-1}$}

\newcommand{\oiii}{[\ion{O}{3}]}
\newcommand{\msunsqpc}{\solmass~pc$^{-2}$}

\begin{document}

\title{Molecular Disks in the Elliptical Galaxies NGC 83 and NGC 2320}
\author{L. M. Young}
\affil{Physics Department, New Mexico Institute of Mining and Technology,
Socorro, NM 87801}
\email{lyoung@physics.nmt.edu}

\begin{abstract}  
The molecular gas in (some) early type galaxies holds important clues to
the history and the future of these galaxies.  
In pursuit of these clues
we have used the BIMA millimeter array to map CO emission in the giant elliptical
galaxies NGC~83 and NGC~2320 and to search for CO emission from the S0
galaxy NGC~5838. 
We also present $V$ and $R$ images of NGC~83 and NGC~2320 which
trace their dust distributions
and enable a search for disky stellar structures.
The molecular gas in NGC~83 is well relaxed, but both CO and dust in
NGC~2320 show asymmetric structures which may be linked to a recent
acquisition of the gas.
However, the specific angular momentum distribution of molecular gas in
NGC~2320 is consistent with that of the stars. 
Internal origin of the gas (stellar mass loss) cannot, therefore, be ruled
out on angular momentum grounds alone.
We also consider the evidence for star formation activity and disk growth
in these two elliptical galaxies.
Radio continuum and FIR fluxes of NGC~83 suggest star formation activity.
NGC~2320 has bright \oiii\ emission, but its large radio/FIR flux ratio and
the mismatch between the
kinematics of CO and \oiii\ suggest that the ionized gas should not
be attributed to star formation.
The origin and future of these two CO-rich early type galaxies are thus
complex, multi-faceted stories.
\end{abstract}


\section{Introduction}

Since the time of IRAS it has been known that a significant number --
perhaps 50\%-- of early
type (E and S0) galaxies have detectable far-IR emission in the 60\micron\ and
100\micron\ bands \citep{knapp89}.
The galaxies which are FIR-bright sometimes also have 
molecular gas; for example,
\citet{knapp96} quote CO detection rates of 20 to 80\%
for ellipticals which are brighter than 1 Jy at 100\micron.
Thus some of the galaxies which are {\it supposed} to be poor in cold gas are
not, which raises a variety of interesting issues.
In spiral galaxies, for example, we are accustomed to thinking of very
intimate connections between the stellar disks and the cold gas.
What about the early type galaxies which are rich in cold gas --- what (if
any) is the connection between their stars and their gas?
Where did this cold gas come from, and how does it evolve?

The CO-rich early type galaxies could have acquired their cold gas from
external sources (including accretion and major mergers) or from internal
stellar mass loss.
The latter source is plausible in the sense that
the rates of mass return to the interstellar medium
over 10 Gyr are more than adequate to account for the 
observed cold gas masses \citep{faber76,ciotti91,bm97}.
After all, the molecular gas masses are still typically much smaller, per unit
optical luminosity, than for spirals \citep{lees91}, so that the CO-rich
early type galaxies are only rich in comparison to other early types.
However, if the molecular gas has an internal origin,
one might expect to find a trend of increasing CO
mass in galaxies of higher optical luminosity.
Such trends are not observed.  Most authors then
suggest that the cold gas in early type galaxies must have been acquired
from outside or as a leftover from a major merger \citep{knapp96, lees91}. 
Therefore, the relatively few early type galaxies which do have
molecular gas could hold vital clues to the evolution of the
early types. 

Beyond simply noting the molecular gas content of early type galaxies, 
observations which resolve the {\it distribution and kinematics} of the gas 
are crucial for understanding its origin and future.
For example, gas which originated in stellar mass loss must have a specific
angular momentum distribution which is consistent with that of the stars.
Such gas may have somewhat less angular momentum than the stars do, but it
should certainly not be misaligned or counterrotating.  In this way the
comparison of stellar and cold gas kinematics can help to clarify the origin
of the gas.
In addition, molecular gas is the raw material for star formation; the
properties of the molecular gas determine where and how star formation may
happen.
The semi-analytic simulations of \citet{khochfar04} suggest that a few tens
of percent of all disky ellipticals could have grown their stellar disks 
out of cold gas accreted from the intergalactic medium.
A better understanding of the molecular gas in early type galaxies could
provide evidence either for or against this disk growth scenario. 

The distribution and kinematics of the molecular gas are also valuable
for mass modelling.
The dissipational nature of the cold gas means that the shapes of the 
gas orbits are much better known than are the shapes of the stellar orbits 
\citep{dezeeuw89, cretton00}.
Gas kinematics can be used to infer the galaxy potential in a way which
is more robust than, or at least complementary to, 
what one can do with stellar kinematics.

There are now around a dozen elliptical or possible elliptical galaxies in the
literature with high quality CO maps which resolve their molecular
distributions \citep{inoue96, young02, okuda05, rupen97, wiklind97}.
(The phrase ``possible ellipticals" is used here because of the well-known
difficulty in separating ellipticals from S0s, especially for photographic
material.
Two merger remnants with $r^{1/4}$ structure are also counted.)
Most have molecular gas in a rotating disk of radius $\sim$ 1
to a few kpc in radius.   These radii are comparable to the effective or
half-light radii $r_e$.
The molecular masses are about \tenup{9} \solmass. 
\citet{young02} shows that such molecular disks could build dynamically
cold stellar disks similar those found in disky ellipticals.
The unusual ellipticals Cen A (NGC~5128) and Perseus A (NGC~1275)
have molecular gas associated with shells \citep{charmandaris00} and H$\alpha$ filaments
\citep{hatch05}, but these are the only ellipticals known to have molecular gas
at such large radii.
Thus, the study of the distribution and kinematics of molecular gas in
early type galaxies is really just beginning.

This paper presents new CO maps which resolve the distribution and
kinematics of molecular gas in NGC 83 and NGC 2320, which are now two of
the most luminous early type galaxies with such maps.
A deep search for CO emission from NGC 5838 is also presented.
Broadband optical images of NGC 83 and NGC 2320 reveal the dust
distributions and optical structure of these galaxies, and we discuss
correlations between dust, CO, and stellar structures.
We also present comparisons of stellar, molecular, and ionized gas
kinematics in NGC 2320; the cumulative specific angular momentum
distributions of molecular gas and stars help to reveal the origin of the
cold gas.

We find some evidence for both an internal origin (stellar mass loss) 
and an external origin of the molecular gas in NGC 83 and NGC 2320.  The
gas/dust disk in NGC 83 is very well relaxed though the gas in NGC 2320
shows some strong asymmetries which should shear away on quite short
timescales.  The specific angular momentum distribution of the molecular
gas in NGC 2320 may be consistent with an internal origin, however.
Evidence for star formation activity and disk growth is also mixed.
The radio/FIR flux ratio in NGC 2320 suggests that the galaxy has relatively
little star formation activity;  it is bright in \oiii\ emission, but the 
kinematics of \oiii\ do not match those of the molecular gas.
In short, it is clearly too early to regard the origin and the future of
the CO-rich early type galaxies as a mystery solved, but the new CO maps of
NGC 83 and NGC 2320 surely provide important clues to an emerging picture.

\section{Selection and properties of the galaxies}

NGC 83, NGC 2320, and NGC 5838 were selected for CO mapping because they are early type 
galaxies with relatively strong CO detections in the single-dish surveys of
\citet{wiklind95} [WCH95] and \citet{knapp96}.
Both of those surveys observed morphological ellipticals
with IRAS 100\micron\ fluxes $S_{100 \mu m} > 1.0$ Jy.
(Many of the morphological classifications have not been checked with modern CCD images.)
WCH95 detected CO emission in 16 of the 29 elliptical 
galaxies they searched.
Their 7 strongest CO detections in the northern hemisphere were mapped
previously \citep{young02}, and NGC 83 and NGC 2320 are two of the next
brightest.  
\citet{knapp96} detected CO emission in 11 of 42 early type galaxies, and 
NGC 5838 has one of the largest fluxes in that sample.

NGC 2320 and NGC 5838 were also selected because of the opportunity to
compare the kinematics of cold gas and stars in early-type galaxies.
Longslit stellar and ionized gas kinematics are available for multiple
position angles in NGC 2320 from the work of \citet{cretton00}.
Furthermore, both NGC 2320 and NGC 5838 have been mapped
in the SAURON survey of early type galaxies \citep{dezeeuw02}, so that full
two-dimensional maps of stellar
line-of-sight velocity profiles, ionized gas kinematics and line strength
indices will be available for the inner parts of these galaxies. 
The claimed CO detection in NGC 5838 has not been
confirmed, but we are still able to compare stellar and gas kinematics for
NGC 2320.

Table \ref{sampletable} gives some properties of the three galaxies
which are studied in this paper.
The available data indicate that NGC 83 and NGC 2320 
are luminous ellipticals whose structure and stellar populations
are typical of their class.
\citet{kuntschner01} show that H$\beta$ and $\rm \langle Fe\rangle$ absorption in NGC
2320 are consistent with solar metallicity and a large luminosity-weighted
mean age. 
NGC 2320 does, however, have the strongest \oiii\ emission in the sample of
\citet{kuntschner01}.
H$\beta$ and Mg$_2$ indices for NGC 83 are also typical of the early
type galaxies in the sample of \citet{trager98}.
WCH95 consider NGC 83 to be a member of a poor group of mostly early type
galaxies and NGC 2320 to be a member of the cluster Abell 569.

NGC~5838 is a well-studied S0 galaxy with a thick nuclear dust ring
\citep{ravindranath01,peletier99}.  
It also contains a nuclear kinematically decoupled region which has the
same colors as the rest of the bulge \citep{falcon-barroso03}.
The large scale structure is discussed at greater length by
\citet{sandage94},
and \citet{michard94} suggest that the outer disk may show faint signs of
a spiral pattern.
In the SAURON sample of early-type galaxies \citep{dezeeuw02}, NGC~5838 is classed as a field
lenticular.

\section{Observations and data analysis}

\subsection{Optical imaging of NGC 83 and NGC 2320}

Since NGC 83 and NGC 2320 are two of the most unusually CO-rich early type
galaxies, and since their morphology has been rather meagerly studied in
the past, it is worthwhile to reinvestigate their structure with high
quality optical images.
Early type galaxies sometimes contain surprises \citep{donzelli03}.
Broadband $V$ and $R$ images of NGC 83 and NGC 2320 were obtained by L.\ van Zee in November 2002
with MiniMo on the WIYN 3.5m telescope under non-photometric
conditions.\footnote{The WIYN Observatory is a joint facility of the
University of Wisconsin-Madison, Indiana University, Yale University, and
the National Optical Astronomy Observatories.}  The total exposure time was 
1260 sec in $V$ and 900 sec in $R$; the pixel scale (after binning) is
0.28\asec\ pix$^{-1}$, the field of view is 10\arcmin, though the galaxies
are not centered in the field, and the
seeing was 1.0\asec\ -- 1.2\asec.
The images were processed in the standard manner for overscan, bias, flat
field and illumination correction, then combined.  Additional data
processing details may be found in \citet{vanzee04}.

For comparison to radio data, sky coordinates were obtained from 6 to 12
stars in the POSS~II digitized sky survey images, and the optical images 
were rotated to have north up.  CO images were regridded to match
the coordinate system in optical images.
Dust images were constructed by first convolving the $V$ images to match the
resolution of the $R$ images and then dividing the two.
After careful masking of stars, galaxy isophotes were fit using
the ELLIPSE task inside the STSDAS package.  Ellipse fits were performed on
both $R$ and $V$ images independently.
These fitted isophotes also give surface brightness profiles which 
were fit to a Sersic function \citep{ciotti99}:
$$  \mu(R) = \mu_0 + \frac{2.5\;b}{\ln 10}(R/R_e)^{1/m}.$$
Sersic fits used the ellipse semimajor axis as the
independent variable (rather than the geometric mean axis).

\subsection{CO data}

NGC 83, NGC 2320 and NGC 5838
were observed in the $^{12}$CO $J=1\rightarrow 0$ line with the 10-element Berkeley-Illinois-Maryland
Association (BIMA) millimeter interferometer at Hat Creek, CA
\citep{welch96}.
These observations were carried out in the C configuration (projected 
baselines 3 to 34 k$\lambda$) between October 2002 and December 2002.
Four tracks of data were obtained for each galaxy with a single pointing
centered on the optical position.
Each observation covered a velocity range of 1300 to 1400 \kms\ centered on
the single-dish CO velocity, and
these data have sensitivity to structures from point sources up 
to objects 60--80\arcsec\ in diameter. 
System temperatures were mostly in the 200--400 K range, with the
exception that three of the tracks on NGC~5838 had system temperatures ranging
from 400 to 600 K.
Table \ref{obstable} summarizes important parameters of these observations.

Reduction of the BIMA data was carried out using standard tasks in the
MIRIAD package \citep{sault95}.
Electrical line length calibration was applied to the tracks on NGC 5838,
which were taken after the array switched from coaxial cable to optical
fiber; for NGC~83 and NGC~2320 it was not needed.
One of the tracks on NGC~5838 was also explicitly corrected for amplitude
decorrelation on longer baselines using data from an atmospheric phase
monitor and  the MIRIAD task {\it uvdecor} \citep{lay99, akeson98, regan01,
wong01}.
The atmospheric decorrelation is estimated using a small interferometer
with a fixed 100 meter baseline which measures the rms path length
difference in the signal from a commercial broadcast satellite.
This corrected track had rms path
lengths in the range of 200 to 300 microns, and the median amplitude
correction factor was 3\%.
The remainder of the tracks were taken in stabler weather and were not
explicitly corrected for decorrelation because normal amplitude calibration
can take out most of the effect \citep{wong01}.

Absolute flux calibration was based on observations of Uranus or Mars.
When suitable planets were not available, the secondary
calibrator 3C454.3 was used as it is usually monitored several times 
per month.
Phase drifts as a function of time were corrected by means of a
nearby calibrator observed every 30 to 40 minutes. 
Gain variations as a function of frequency were corrected by the online
passband calibration system;
inspection of data for 3C273 indicate that residual
passband variations are on the order of 10\% or less in amplitude and 2\deg\ in
phase across the entire band.

The calibrated visibility data were weighted by the inverse square of the
system temperature and the inverse square of the amplitude 
decorrelation correction factor (if used), then Fourier transformed.
NGC 83 and NGC 2320 showed some line emission;
the dirty images were lightly deconvolved with the Clark clean algorithm,
as appropriate for these compact, rather low signal-to-noise detections.
(No continuum subtraction was needed for these galaxies.)
Integrated intensity and velocity field maps were produced by the masking 
method: 
the deconvolved image cube was smoothed along both spatial and velocity 
axes, and the smoothed cube was clipped at about 2.5$\sigma$ in absolute
value.
The clipped version of the cube was used as a mask to define a
three-dimensional volume in which the emission is integrated over velocity
\citep{wong01,regan01}.
Continuum images were made by averaging all of the line-free channels in
the final spectral line cubes.
Table \ref{obstable} gives 3$\sigma$ limits for point source continuum
emission at the centers of the galaxies;
Table \ref{imgtable} gives beam size and sensitivity information for the
final spectral line cubes.

\section{Results}

\subsection{NGC 83: optical morphology}\label{83optical}

\begin{figure}
\includegraphics[scale=0.8]{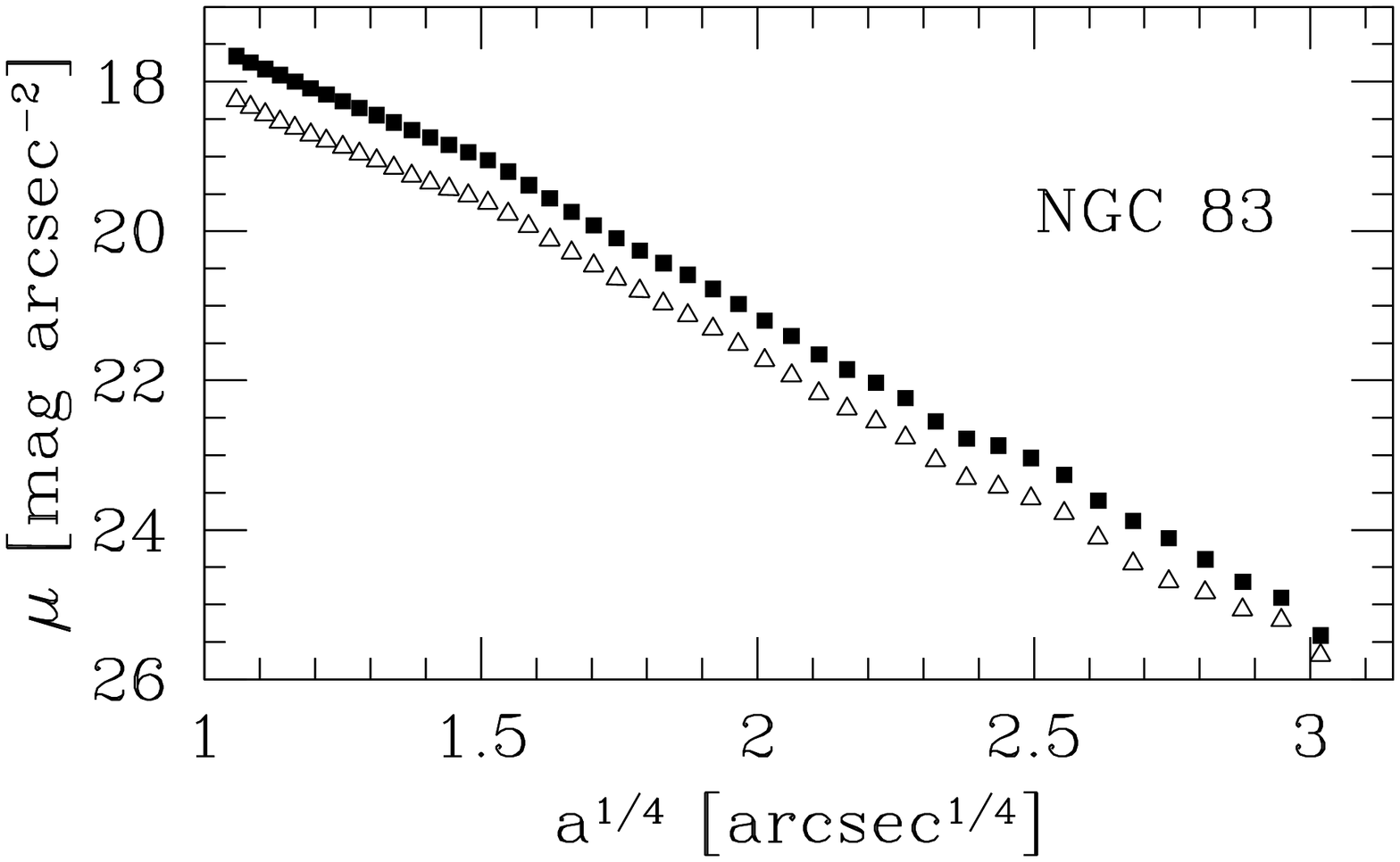}
\caption{Surface brightness profiles of NGC 83.
$R$ is in filled squares and $V$ is in open triangles.
The break in slope at $a^{1/4}\sim 1.5$~arcsec$^{1/4}$ is due to the dust
disk.
Approximate flux calibration for these profiles is derived by matching
the
photometry in circular apertures to data in the hyperLEDA database; it is
estimated to be accurate to $\pm$ 0.05 mag.
Corrections for Galactic extinction are also applied, amounting to 0.23
magnitudes in $V$ and 0.185 mag in $R$ \citep{schlegel98}.
\label{83profile}
}
\end{figure}

Over a radial range of 6\asec\ to at least 80\asec\ 
(exterior to a dusty region described below), 
the surface brightness profile of NGC 83 is very well described by a $r^{1/4}$
profile (Figure \ref{83profile}).
In fact, Sersic fits give
fitted values of the shape parameter $m$ between 4.1 and 5.1
for reasonable radius ranges.
A value of $m=4$ would correspond exactly to a $r^{1/4}$ profile.
Formal uncertainties in the fitted value of $m$ are on the order of 0.07,
but it is clear that the range of values indicated above is a better
indicator of the true uncertainty.

\begin{figure}
\includegraphics[scale=0.8]{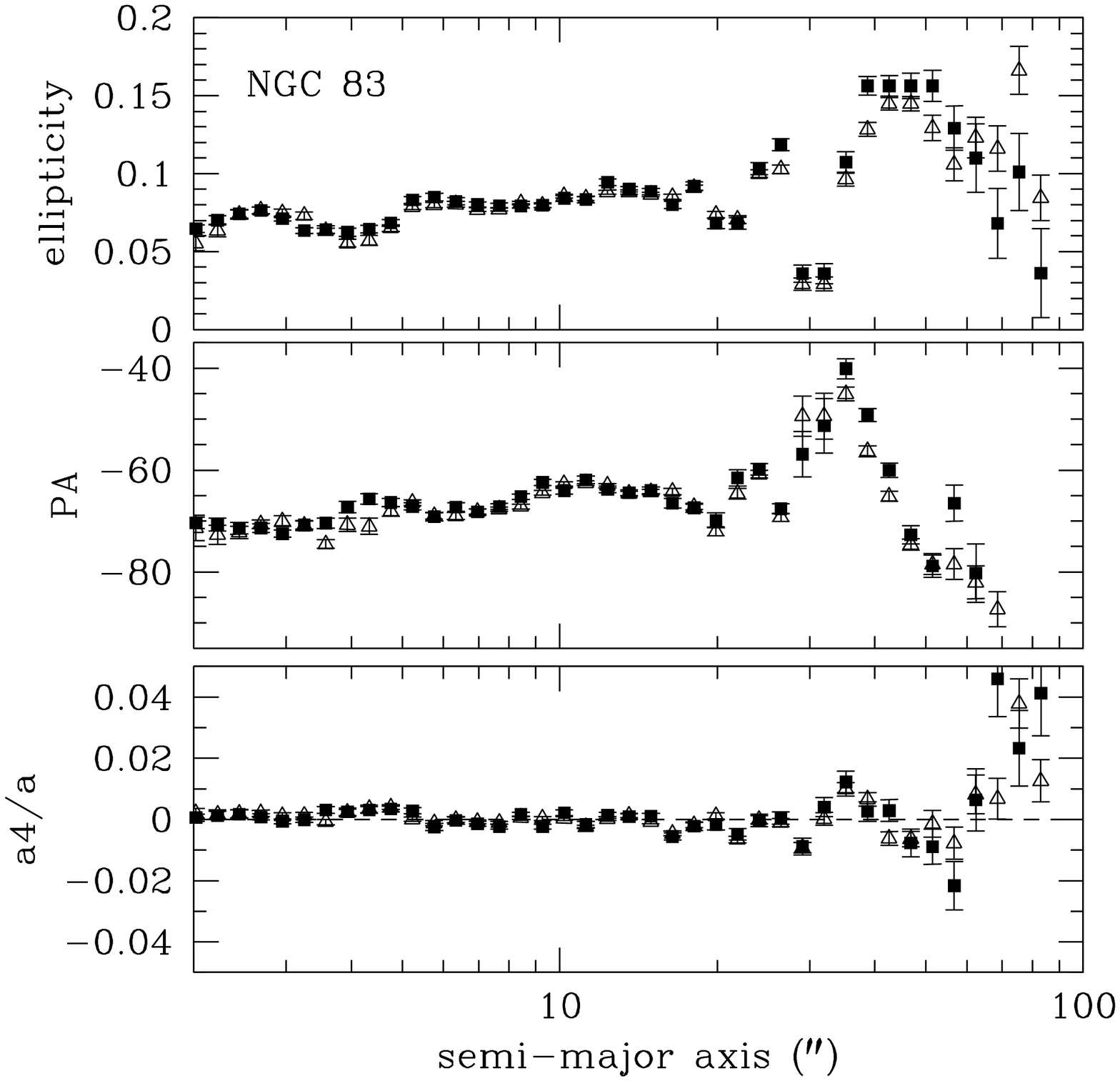}
\caption{Optical morphology of NGC 83.
The isophote ellipticity, position angle, and deviations from pure ellipses 
($a_4/a$) are indicated for the $R$ image (filled squares) and the $V$
image (open triangles).
Fitted parameters at $a \leq 5''$ are affected by dust.
\label{83epa}
}
\end{figure}

The galaxy's position angle remains constant at $-66\deg$ 
over 6\asec\ to 30\asec\ (Figure \ref{83epa}).
The ellipticity of the isophotes is also flat at a mean value of $\epsilon =
0.085$ over this radius range, and the isophotes show
no significant deviation from pure ellipses.
From 30\asec\ outwards, the structure of the galaxy is somewhat more
complicated; a slight ``wing" or knee in the surface brightness at 
30-42\asec\ (and probably some scattered light from a trio of
bright stars) are responsible for excursions in ellipticity, position
angle, and the Fourier components a3 and a4.
Thus, the surface brightness profile of NGC 83 is entirely consistent with 
its classification as an elliptical, and the interior is
neither boxy nor disky with no position angle twist.

\begin{figure}
\includegraphics[scale=0.5]{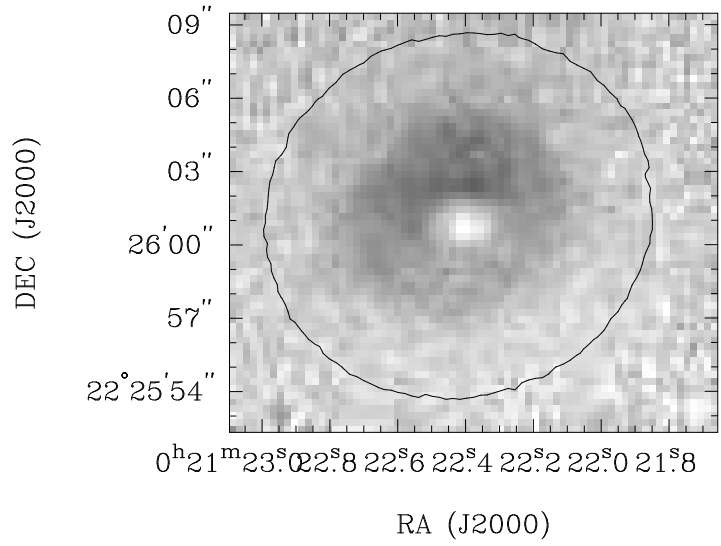}
\caption{$V-R$ color in NGC 83.  Darker pixels have redder $V-R$ colors.
An isophote from the $R$ image is superposed.
\label{n83dust}
}
\end{figure}

The $V-R$ dust image (Figure \ref{n83dust})
shows a very regular, relaxed disk of semimajor axis $\sim$ 5\asec\ which is well aligned with the optical
major axis.  The northeast side of the disk shows somewhat stronger 
reddening and
must be the near side.  The axis ratio of the dust disk is estimated
in the range $b/a = 0.8~{\rm to}~0.9$, implying that the inclination of
the disk is between 26\deg\ and 37\deg.  The stellar body of the galaxy is
just marginally rounder than the dust disk, and 
there is no evidence for dust lanes beyond this central disk.
The color image also shows that the nucleus is bluer than the
rest of the galaxy.

\subsection{NGC 2320: optical morphology}\label{2320optical}

\begin{figure}
\includegraphics[scale=0.8]{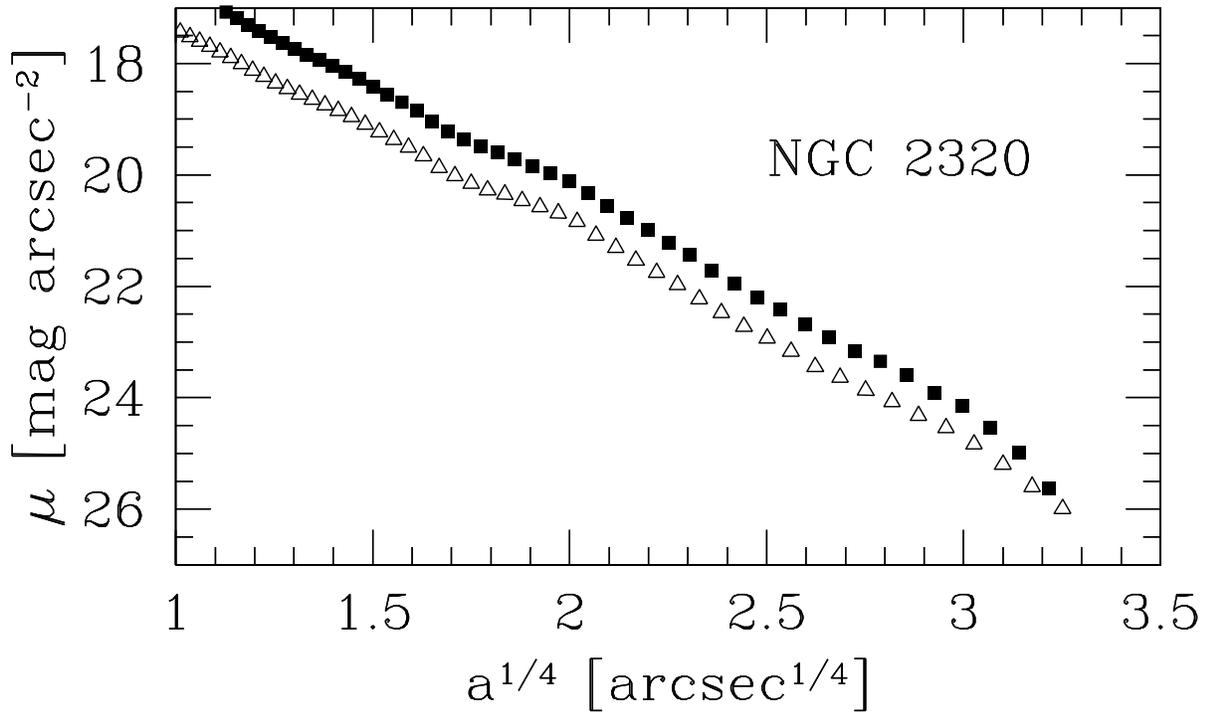}
\caption{Surface brightness profiles of NGC 2320.
Dust extinction is visible at radii $\leq 12''$.
As for Figure \ref{83profile}, the surface brightness profiles were roughly
calibrated with circular aperture photometry from the hyperLEDA database.
Galactic extinctions of 0.226 mag ($V$; triangles) and 0.182 mag ($R$;
squares) were applied \citep{schlegel98}.
\label{2320profile}
}
\end{figure}

\begin{figure}
\includegraphics[scale=0.8]{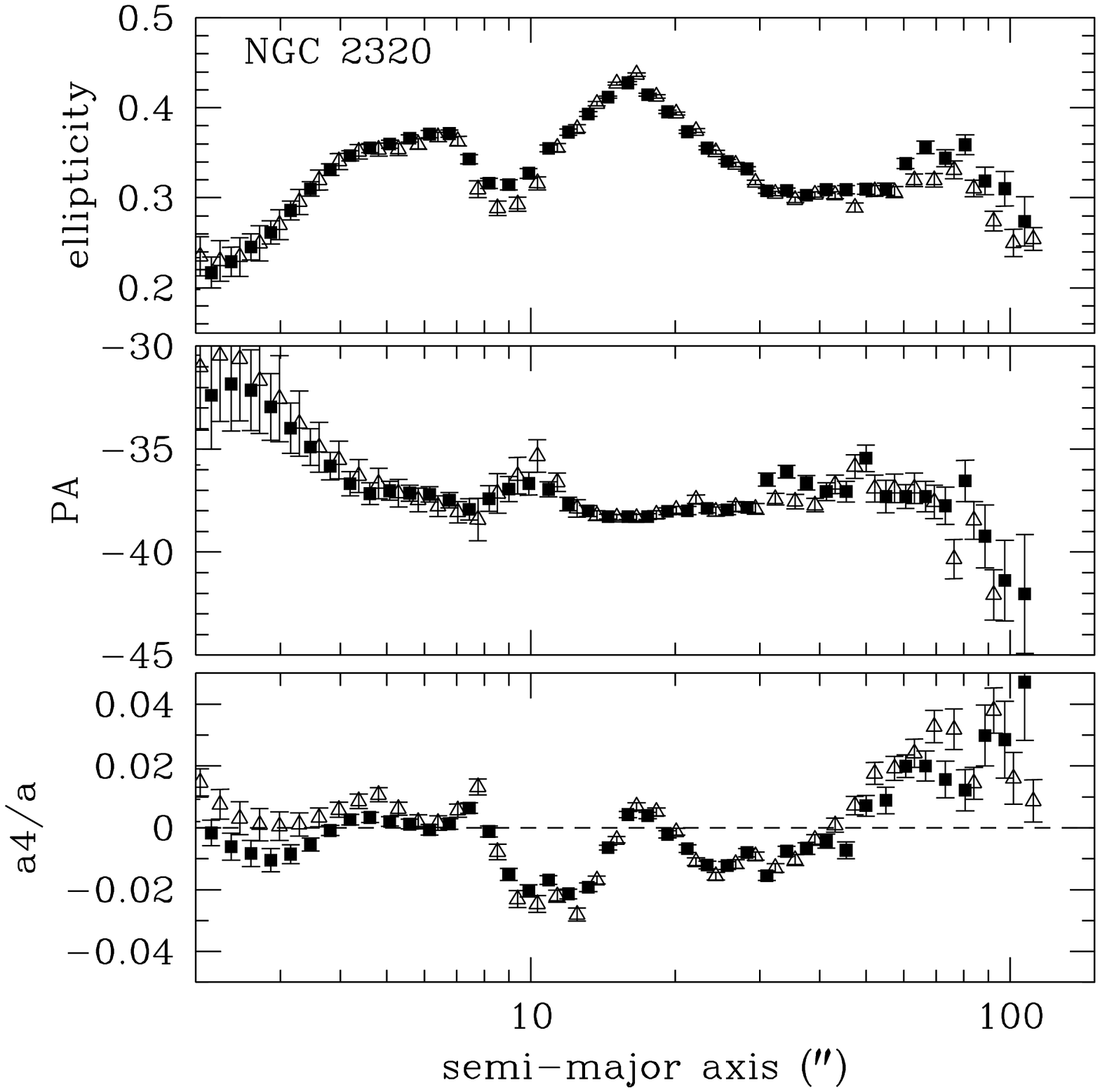}
\caption{Optical morphology of NGC 2320.
The isophote ellipticity, position angle, and deviations from pure ellipses 
($a_4/a$) are indicated for the $R$ image (filled squares) and the $V$
image (open triangles).
Fitted parameters at $a \leq 12''$ are affected by dust.
\label{2320epa}
}
\end{figure}

The optical morphology of NGC 2320 is also consistent with its
classification as elliptical, though its structure is more complicated than
that of NGC 83.
Over semimajor axes 14\asec\ to
90\asec\ (there is dust absorption interior to 12\asec), 
its surface brightness is well described by a $r^{1/4}$  profile
(Figure \ref{2320profile}).
The fitted Sersic shape parameters fall in the range $m=4.0$ to 5.2 for
reasonable radius ranges.
The position angle is consistent with a constant value (no twists) of 
$-38\deg$ from 14\asec\ to 70\asec\ (Figure \ref{2320epa}).
A bright annulus of semimajor axis 15\asec\ 
shows up as a small bump in the surface brightness profile and a prominent
peak in ellipticity.  Just interior to the bright annulus 
the isophotes are strongly boxy ($a_4/a \sim -0.02$ near
semimajor axis of 10\asec) but the bright annulus itself shows $a_4/a
\sim 0$.  Beyond 30\asec\ we find a constant ellipticity $\epsilon =
0.3$ and a gradual transition from $a_4/a = -0.01$ (boxy) at 30\asec\ to 
$a_4/a \sim +0.02$ (disky) at 70\asec.
These results agree reasonably well with those of \citet{nieto91} if
the scaling on the semimajor axis of their Figure 2.14 is off by about a factor of
two, though
we do not confirm the position angle drift claimed by those authors.

\begin{figure}
\includegraphics[scale=0.5]{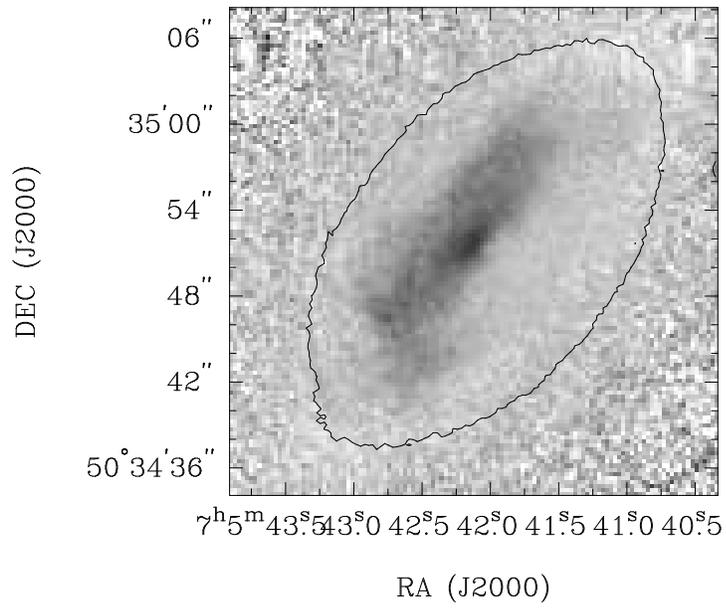}
\caption{$V-R$ color in NGC 2320.  Darker pixels have redder $V-R$ colors.
An isophote from the $R$ image is superposed.
\label{n2320dust}
}
\end{figure}

The $V-R$ image (Figure \ref{n2320dust}) shows significant reddening by
dust within 12\asec\ of the nucleus.  The northeast side of the disk is the
near side.
The bilateral symmetry
of the molecular gas distribution (Section \ref{2320CO}) makes it clear
that the far (southwest) side of the disk is also present, although it is not obvious 
from the reddening alone. 
The dust is also reflection-symmetric about the minor axis out to a
semimajor axis of 10\asec; beyond 10\asec\ there is some ``extra"
absorption southeast of the nucleus.  
The symmetric part of the dust disk has an axis ratio
$b/a = 0.58$ which corresponds to an inclination of 55\deg.
\citet{cretton00} also see this dust feature in residuals to a 
multi-Gaussian fit of their V-band image and estimate its axis ratio to be
0.5 ($i =$ 60\deg) or 0.32 ($i =$ 71\deg) after unsharp masking.
An inclination of 60\deg\ is therefore adopted for the gas and dust disk.

\subsection{CO in NGC 5838}

NGC 5838 is not detected in CO emission.
The \htoo\ column density limits in Table \ref{imgtable} correspond to a mass
limit $M(H_2) < 5\times 10^{6}$ \solmass\ for an unresolved source at the center of
the field of view.
(The adopted \htooco\ conversion factor is discussed in Section
\ref{fluxes}.)
This upper limit is well below the expected CO mass based on the
work of \citet{knapp96}, whose
CO $J=2\rightarrow 1$ spectrum shows a peak
brightness temperature ($T_{mb}$) of about 10 mK and an rms noise level of
5.4 mK per 10 \kms\ channel.  The shape is reminiscent
of a two-horned spectrum from a rotating disk, with the optical velocity
roughly midway between two peaks.  
Our BIMA observations achieved an rms noise level of 17 \mjb\ = 14 mK in a data
cube with 40 \kms\ channels and an 11$\times$10\asec\ beam.  
If the $J=2\rightarrow 1$ and $J=1\rightarrow 0$ transitions have the same
excitation temperature \citep{lees91} and if the 
CO emission were still unresolved in the smaller BIMA
beam, its brightness temperature would rise to 90 mK.
We would therefore have 
expected to detect it at a signal-to-noise ratio of 6 in three or four
independent channels on each ``horn".  Clearly, we did not detect
emission from NGC 5838 at those levels.  

Of course, if the emission in NGC~5838 is extended on the
30\asec\ scales probed by CSO it would still be 10 mK bright at the higher
resolution and our BIMA data would not detect it,
but the dust distribution 
suggests that this is probably not the case.
The dust disk which shows up prominently in the HST image of 
\citet{ravindranath01} has a semimajor axis $<$ 4\asec.
Thus, if the CO were associated with dust as it is in NGC 83 and NGC 2320
(Sections \ref{83CO} and \ref{2320CO}) it would be well matched to the BIMA
beam size and bright enough to detect with the interferometer.

Our present nondetection of CO $J = 1 \rightarrow 0$ emission from NGC 5838 is also
consistent with a recent nondetection at the IRAM 30m telescope
\citep{combes05}.  Thus, we infer that 
the CSO spectrum might simply show
an unfortunate baseline ripple with an amplitude of about 3 mK (smaller
than the rms noise level of those data) and a period of 230 MHz (300 \kms).

\subsection{NGC 83 and NGC 2320: CO fluxes and H$_2$ masses}\label{fluxes}

Total CO fluxes for NGC 83 and NGC 2320 were measured from the 
integrated intensity images; their uncertainties are probably 15\%,
with approximately equal contributions coming from the absolute calibration
and from uncertainties in choosing the spatial region to be summed.
\htoo\ masses 
are calculated using the distances in Table \ref{sampletable} and a
``standard" \htooco\ conversion factor of 3.0\e{20} \convunits\
as in WCH95.  With this conversion
factor, \htoo\ masses  are related to CO fluxes $S_{CO}$ by
$ M(H_2) = (1.22\times 10^4\: M_\odot)\: D^2\: S_{CO}$
where $D$ is the distance in Mpc and $S_{CO}$ is the CO flux in \jykms.
No correction has been made for the presence of helium except where noted.

\begin{figure}
\includegraphics[scale=0.7]{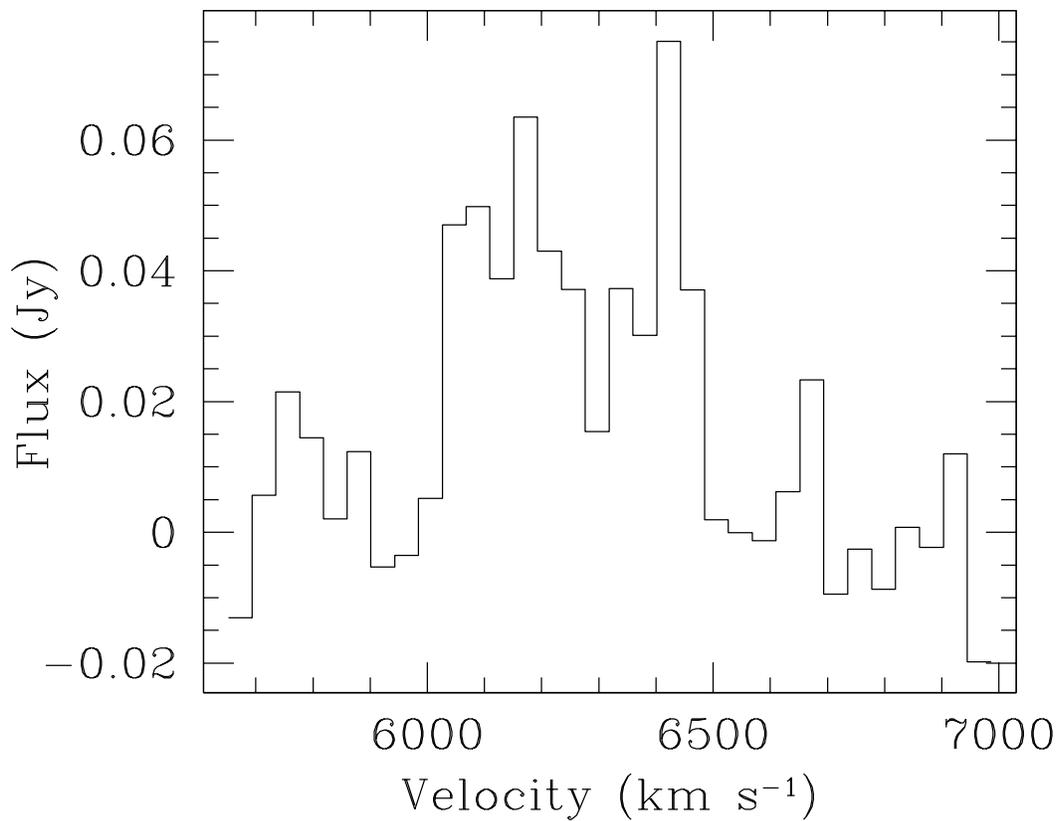}
\caption{CO spectrum of NGC 83.  
The spectrum was constructed by first using the integrated intensity
image (Figure \ref{n83stars}) to define an irregular mask region
within which the emission is located.  The intensity was integrated over
the same spatial region for every channel, so the noise in the line-free
regions of the spectrum should be indicative of the noise on the line as
well.
\label{n83spect}
}
\end{figure}

The interferometric CO flux for NGC 83, 
22\error 3 \jykms, agrees very well with the single dish flux 
measured by WCH95.
The IRAM 30m telescope detected a total flux of 4.7 K~\kms\
(on the $T_{mb}$ scale), 
which corresponds to 21.8 \jykms\ at 4.64 Jy/K \citep{guelin95}.
The uncertainty in the single dish CO flux is roughly 10\% for absolute
calibration, 10\% for statistical uncertainty, and another 5\% from
estimating the baseline level--- in short, the two measurements are
consistent within their errors.
The molecular mass of NGC~83 is therefore (2.0\error 0.3)\e{9} \solmass.
The BIMA CO spectrum (Figure \ref{n83spect}) also shows close agreement
with the center velocity and width found by WCH95 and with the optical
velocity from \citet{huchra99}.

\begin{figure}
\includegraphics[scale=0.7]{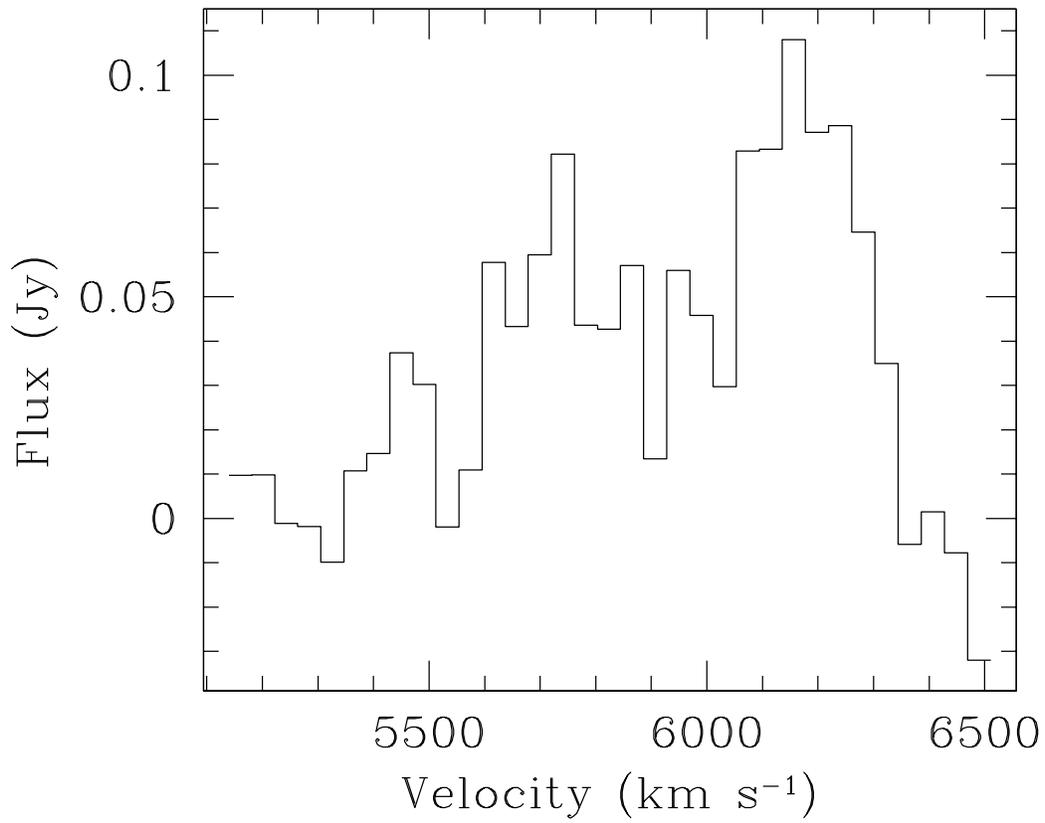}
\caption{CO spectrum of NGC 2320.  The spectrum was constructed in the same
way as Figure \ref{n83spect}.
\label{n2320spect}
}
\end{figure}

Our new CO flux for NGC 2320, 56 \error 8 \jykms, is more than 
five times larger than the total flux given in WCH95 (10 \jykms).
However, it is clear that the discrepancy is caused by the galaxy's large line
width.  Our BIMA data show emission over a range of $\approx$  830
\kms\ (Section \ref{2320CO}).  WCH95 claimed a line width of only 320 \kms; in fact, they only
detected one end of the line, the full 830 \kms\ line being difficult to
distinguish from low level baseline variations in a single dish spectrum.
The total molecular mass of NGC~2320 is thus (4.3 \error 0.6)\e{9}
\solmass.
Our new systemic velocity of 5886 \error 20 \kms\ 
is also in
better agreement with the optical velocity measurements of 5944\error 15 \kms\
\citep{smith00} and 5725\error 60 \kms\ \citep{RC3}.  Figure \ref{n2320spect}
shows the integrated CO spectrum of NGC 2320.

\subsection{NGC 83: CO distribution and kinematics}
\label{83CO}

\begin{figure}
\includegraphics[scale=0.7]{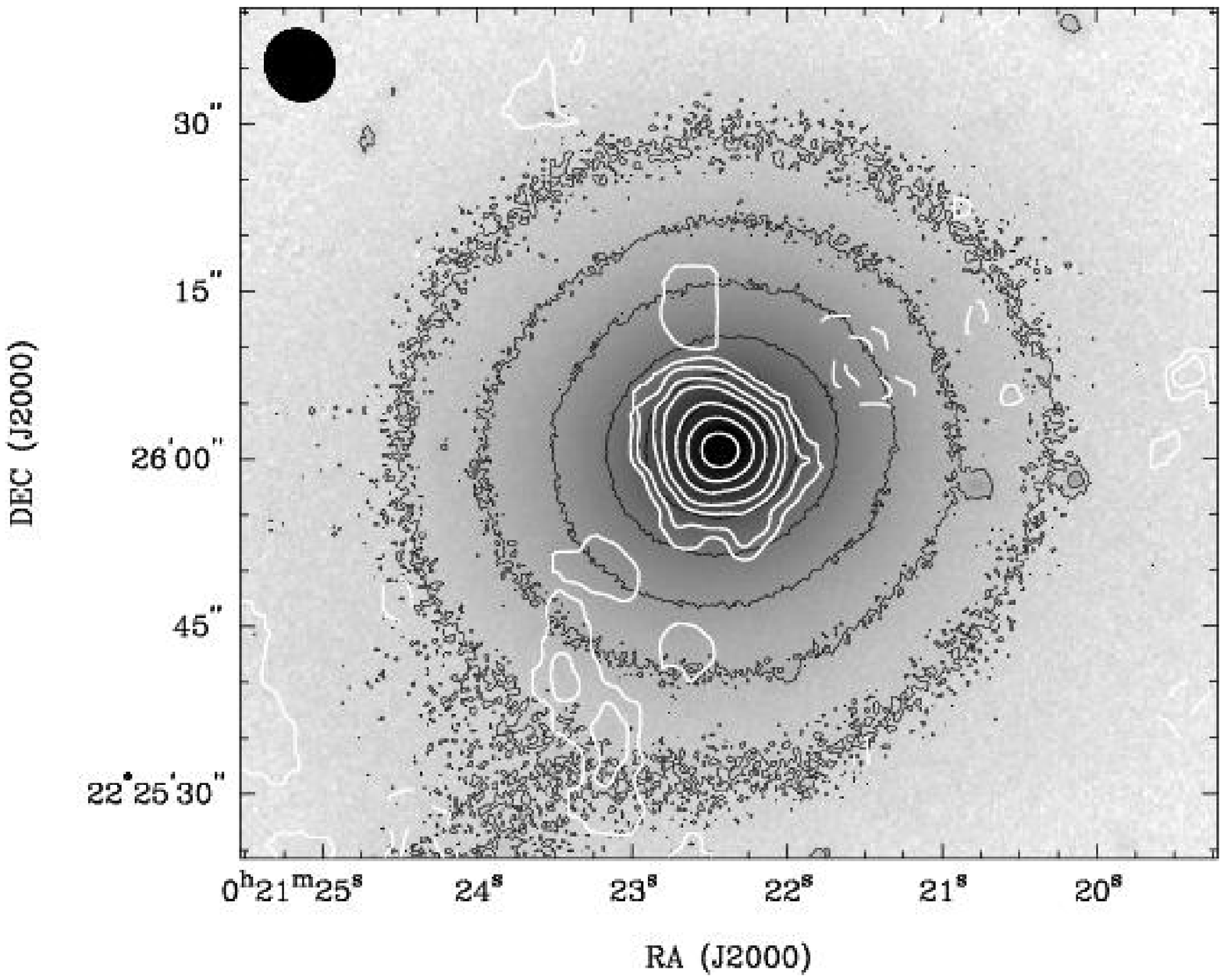}
\caption{Optical and CO images of NGC~83.  The greyscale is the WIYN 3.5m
$R$ image;
black contours are optical isophotes separated by a factor of 2.
White contours are the integrated CO intensity map.
CO contour levels are $-$10, $-$5, 5, 10, 20, 30, 50, 70, and 90 percent
of the peak (12.7 \jybks, or 8.8\e{21} \persqcm).
\label{n83stars}
}
\end{figure}

\begin{figure}
\includegraphics[scale=0.75,clip]{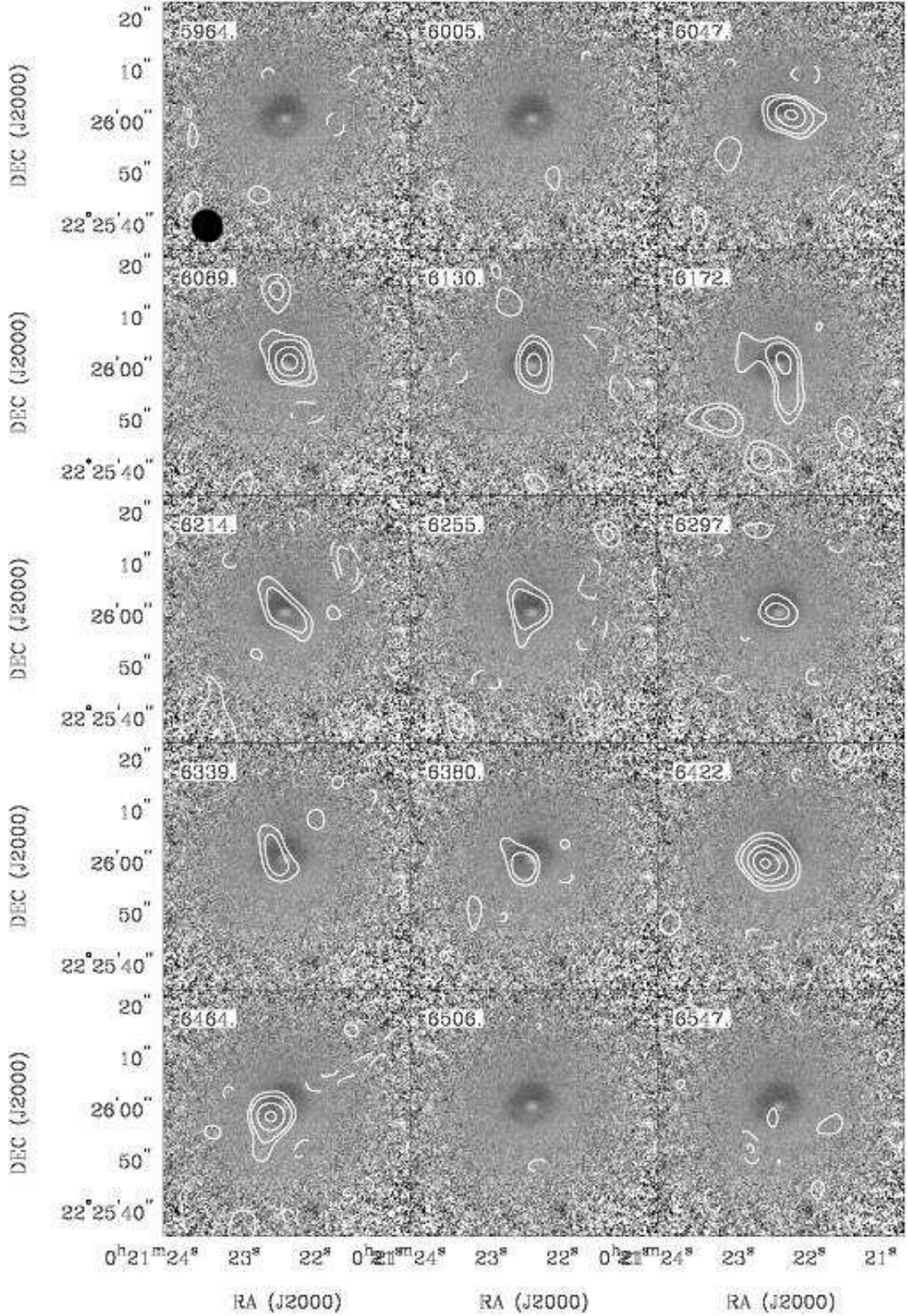}
\caption{Individual channel maps showing CO emission from NGC~83.
Contour levels are $-$3, $-$2, 2, 3, 5, 7, and 9 times 
5.8 \mjb\ =  1$\sigma$.
The greyscale is the $V-R$ image.
The velocity of each channel (in
\kms) is indicated in the upper left corner, and the beam size is shown
in the bottom left corner of the first channel.
The peak CO intensity at 6.6\asec~$\times$~6.0\asec\ resolution 
is 55 \mjb\ = 130 mK.
\label{n83channels}
}
\end{figure}

The distribution and kinematics of CO in NGC 83 are (as best we can tell
for this rather limited resolution) entirely consistent with a symmetric,
relaxed disk in dynamical equilibrium.
A fit of a two-dimensional Gaussian to the integrated intensity image of
Figure \ref{n83stars} gives a deconvolved FWHM of 
5.3$\times$4.4\asec\ at a position angle of 47\deg\ (i.e.\ elongated on the
optical {\it minor} axis), but since that deconvolved size is so similar to
the angular resolution it is more instructive to look at the distribution
in the individual channel maps (Figure \ref{n83channels}).
At the extreme channels (6047 \kms\ and 6464 \kms)
the centroids of the emission are displaced towards opposite ends of the
major axis of the dust disk, and in the interior channels the
emission is elongated along the minor axis of the dust disk.
Thus, the kinematic major axis of the CO is clearly aligned with the major
axis of the dust disk and with the galaxy itself.

CO emission does not appear to extend beyond the dust disk, and indeed the
evidence suggests that the CO emission is exactly coincident with the dust
disk.
The dust disk has a major axis diameter of about 10\asec; thus, the CO
radial extent along the major axis is most likely 5\asec.  
\citet{burstein87} quote an effective radius $r_e$ = 27\asec, so 
the CO/dust disk extends to $0.19\;r_e$.

\begin{figure}
\includegraphics[scale=0.7,bb=26 106 574 669]{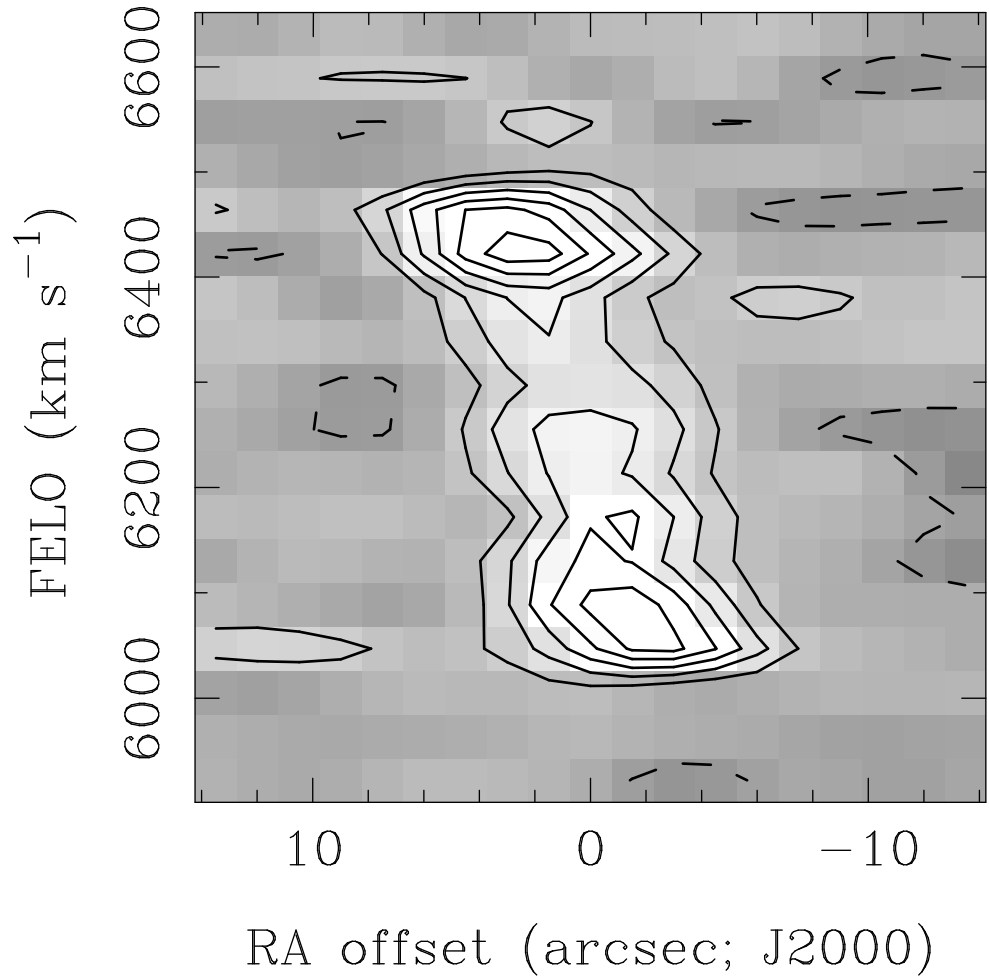}
\caption{Major axis position-velocity diagram for CO in NGC~83.
Contours indicate the intensity at $-15$, 15, 30, 45, 60, 75, and 90
percent of the peak (51 \mjb).
\label{n83pv}
}
\end{figure}

The kinematic center and inclination cannot be determined from the CO data
alone, but the direction of the kinematic major axis and the product
$V\sin(i)$ are well constrained.
A fit of the CO velocity field (intensity-weighted mean velocity) 
with a solid body rotation curve gives the
kinematic position angle to be $-64$ \error 1\deg\ to the
approaching major axis.  
(This formal error estimate is certainly smaller than the true uncertainty,
due to the relatively poor resolution.)
Thus, the kinematic position angle agrees well with the 
value adopted for the optical major axis, $-$66\deg.
A major axis position-velocity slice is shown in Figure \ref{n83pv}.
The circular velocity $V_{circ} \sin(i)$ is estimated from the 
extreme channels which show emission (Figures \ref{n83spect} and
\ref{n83channels}). Since the global profile's edges are sharp compared
to the 40 \kms\ channel width, we can very roughly account for instrumental
broadening and local dispersion by taking the full width to be the
difference in center velocities of those end channels.
Such corrections are discussed in greater detail by \citet{lavezzi98b}, for
example.
Thus the full width of the CO emission is
417 \kms \error 20 \kms, or  $V_{circ} \sin(i)$ = 209 \kms.

Dynamical mass estimates for the center of NGC~83 are in agreement with
expectations.
If the inclination of the dust disk (26\deg\ to 37\deg) is also adopted for
the gas disk, we estimate $V_{circ}$ to be between 350 \kms\ and 480 \kms\ at
5\asec\ = 2.1 kpc.  (The true maximum circular velocity might, of course,
be larger than this if the turnover in the rotation curve is beyond
the edge of the molecular disk.)
The orbital timescale at the outer edge of the CO disk is then 3\e{7} --
4\e{7} yr.
These velocities imply dynamical masses 
between 5.9\e{10} \solmass\ and 1.1\e{11} \solmass\ interior to
2.1 kpc, or $M_{gas}/M_{dyn} = 0.024 - 0.046$ including a factor of 1.36 for
helium.
Approximately 10\%
of the galaxy's light should originate at $r\leq 0.19\:
r_e$; for example, in a \citet{dehnen93} model with $\gamma = 3/2$, 
8.4\% of the stellar mass is interior to $0.19\: r_e$.
NGC~83's absolute magnitude $M_V = -22.1$ then implies that the stellar
luminosity interior to $0.19\:r_e$ is $L_V = 5.1\times 10^{9}$ \solum, and
in that region $(M/L)_V$ = 11 -- 22.  This mass-to-light ratio is very similar to
what \citet{cretton00} found for the center of NGC~2320.

\subsection{NGC 2320: CO distribution and kinematics}
\label{2320CO}

\subsubsection{The CO disk}

\begin{figure}
\includegraphics[scale=0.7]{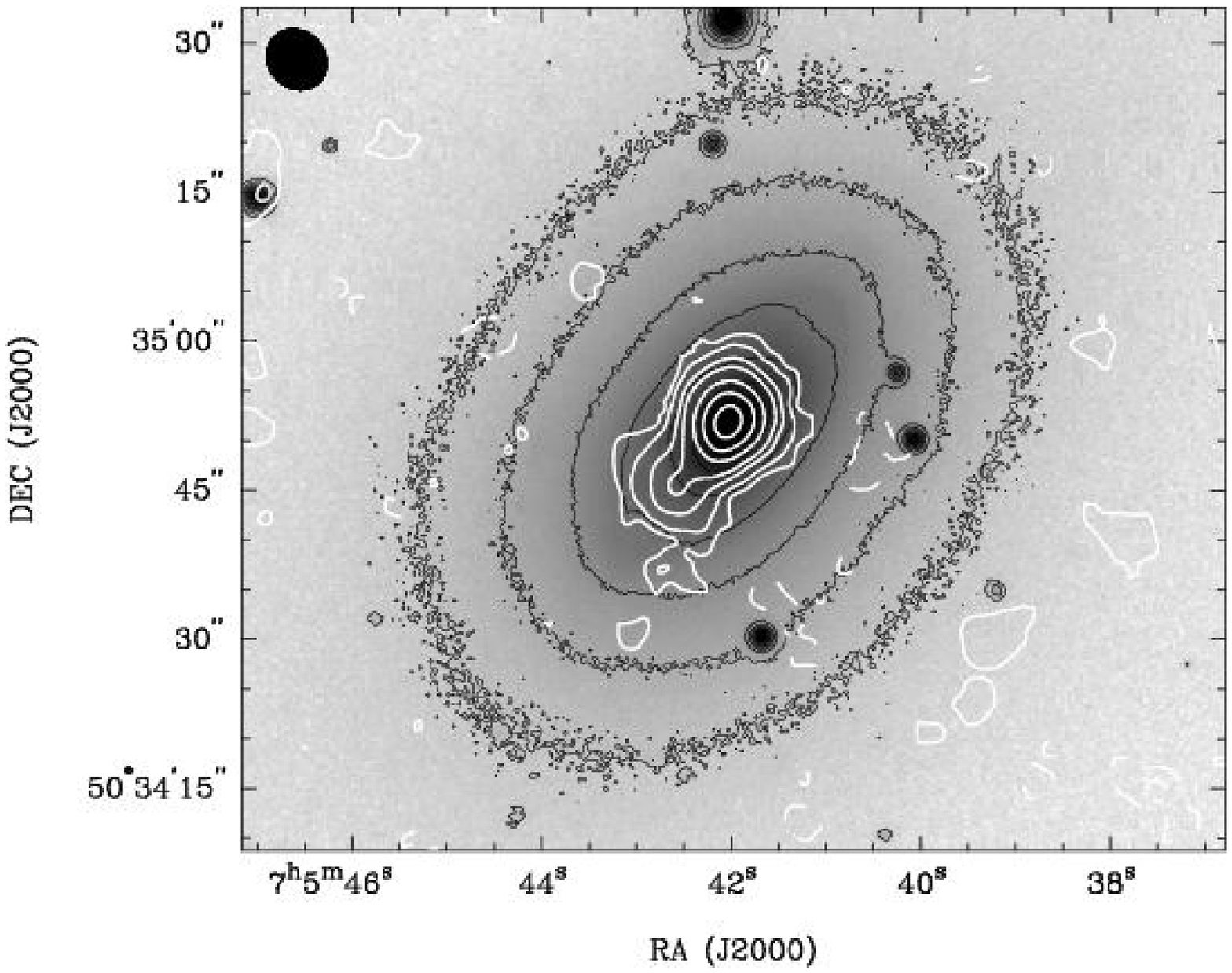}
\caption{Optical and CO images of NGC~2320.  The greyscale is the WIYN 3.5m $R$
image;
black contours are optical isophotes separated by a factor of 2.
White contours are the integrated CO intensity map.
CO contour levels are $-$10, $-$5, 5, 10, 20, 30, 50, 70, and 90 percent
of the peak (30.4 \jybks, or 2.3\e{22} \persqcm).
\label{n2320stars}
}
\end{figure}

\begin{figure}
\includegraphics[scale=0.75,clip]{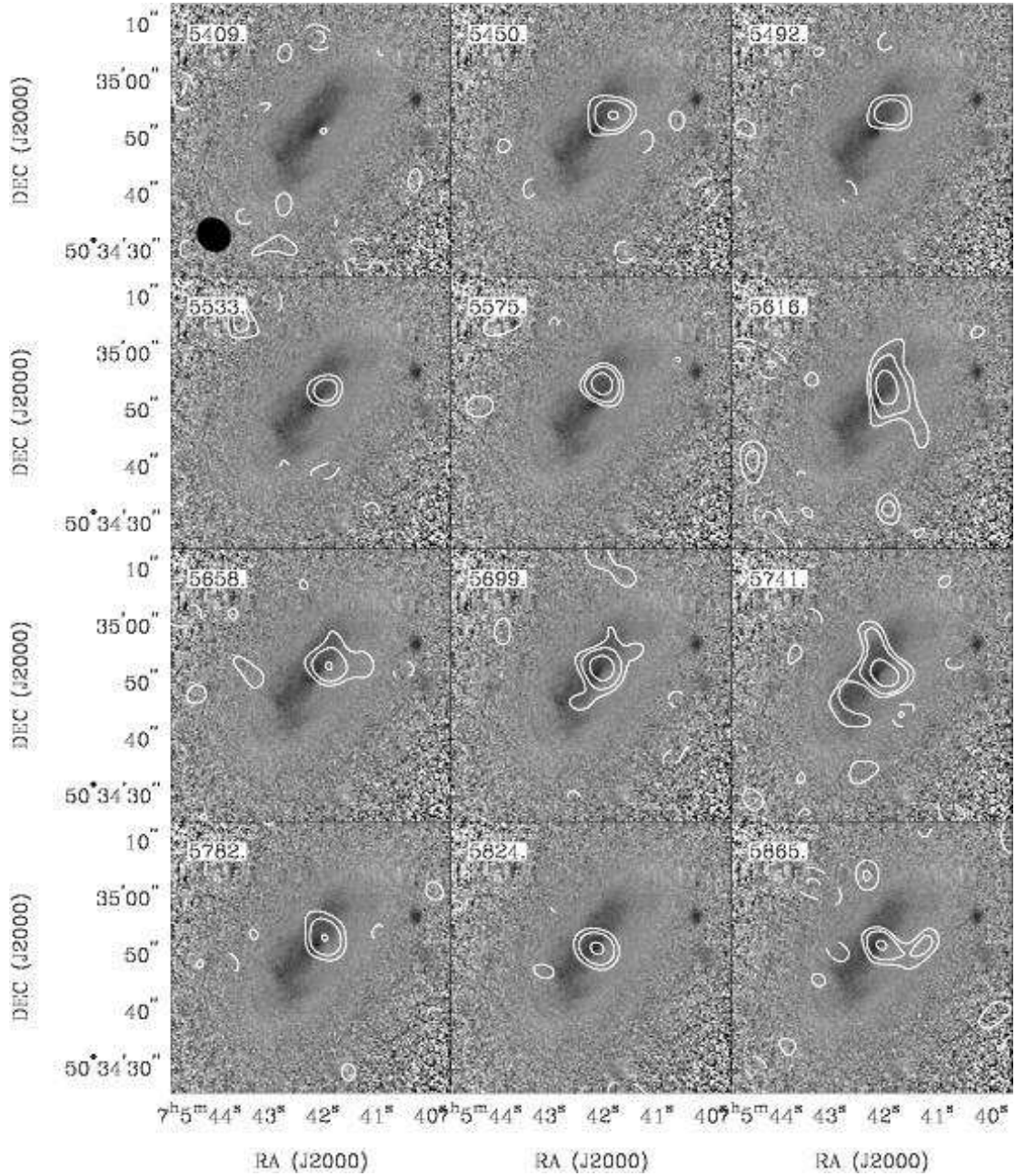}
\figurenum{13}
\caption{Individual channel maps showing CO emission from NGC~2320.
Symbols and contour intervals are the same as for Figure \ref{n83channels}
but the multiplicative unit is 1$\sigma$ = 7.5 \mjb.
The greyscale is the $V-R$ image.
The peak CO intensity at 6.4\asec~$\times$~5.6\asec\ resolution is 61 \mjb\
= 150 mK.
\label{n2320channelsa}
}
\end{figure}

\begin{figure}
\includegraphics[scale=0.75,clip]{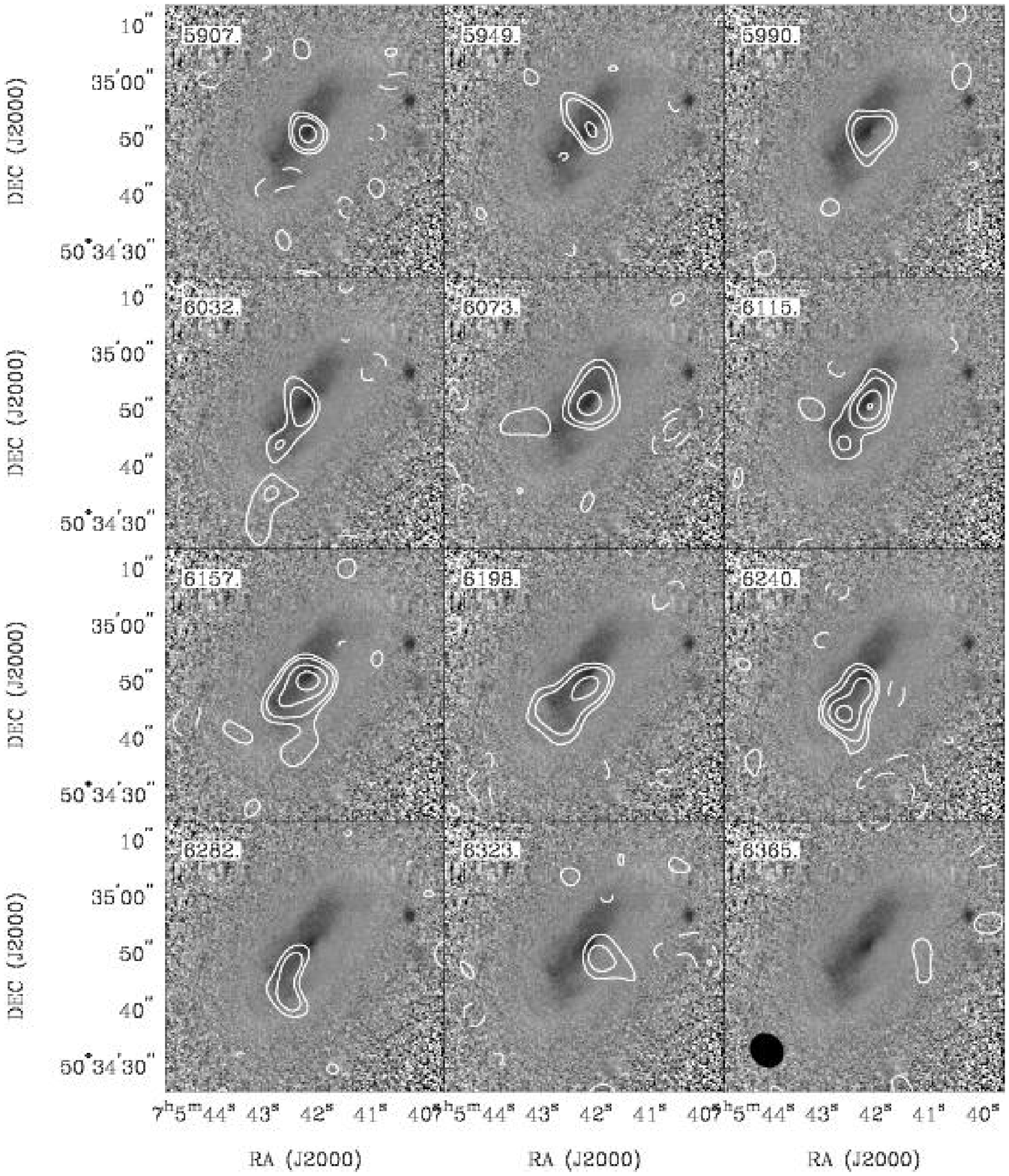}
\figurenum{13}
\caption{Channel maps for NGC~2320, continued from Figure
\ref{n2320channelsa}.
\label{n2320channelsb}
}
\end{figure}

In contrast to the symmetry and relaxed appearance of the CO in NGC 83, the
molecular gas in NGC 2320 is quite asymmetric.
The spectrum in Figure \ref{n2320spect} shows that the flux density is
twice as large on the high velocity side of the line as on the low velocity
side.
From the integrated intensity image (Figure \ref{n2320stars})
it is evident that the ``extra" gas on the high
velocity side is extended along the optical major axis toward the
southeast; it contributes roughly 25\% of the total CO flux of the galaxy, 
and it is coincident with the irregular dust lane just beyond the southeast
end of the dust/gas disk (Figure \ref{n2320dust}).
This extended CO emission is prominent in the channel maps (Figures
\ref{n2320channelsa} and \ref{n2320channelsb})
at velocities $\sim$~6240 \kms, 10\asec\
southeast of the galaxy center.  
Channels at 5450 \kms\ and 6282 \kms\ show
that the CO emission extends to a radius of 4\asec\ (0.1~$r_e$) on the
northwest side of the nucleus and 10\asec\ (0.2~$r_e$) on the southeast
side.

As was the case for NGC 83, the modest resolution does not allow us to determine the
inclination of the gas disk independently from its rotation velocity.
However, tilted-ring fits to the velocity
field do constrain the major axis position angle to be
$-36$\deg \error 4\deg, where the quoted 
uncertainty is estimated from the dispersion among fits with slightly
different radial ranges, initial conditions and weight schemes.
Again, the kinematic major axis is consistent with the optical
morphological major axis ($-38\deg$) though misalignments on
the order of a few degrees cannot be ruled out.

Figure \ref{n2320pv} shows a major axis position-velocity diagram for the
CO emission from NGC 2320.
The full width of the CO line is measured from the extreme channels which
show emission, approximately corrected for turbulent and instrumental
broadening as described in Section \ref{83CO}; it is 830 \kms \error 40
\kms, so $V \sin(i)$ = 415 \kms.
Adopting an inclination of 60\deg\ from the dust disk (Section
\ref{2320optical} and \citet{cretton00}) gives an estimated circular
velocity 480 \kms\ at radii 4--10\asec.
The dynamical mass within 10\asec\ (3.8 kpc) is 2.0\e{11} \solmass,
so $M_{gas}/M_{dyn} = 0.030$ when helium is included.

\subsubsection{Molecular gas vs.\ ionized gas}

Figure \ref{n2320pv} also compares 
CO kinematics in NGC 2320 to stellar and ionized gas 
kinematics from the longslit spectroscopy of \citet{cretton00}.
Those authors used the kinematics of \oiii, corrected for
asymmetric drift, to derive a circular velocity curve for NGC 2320.
Figure \ref{n2320pv} shows that the total width of the CO line is in very
close agreement with the amplitude of the circular velocity curve (assuming
the inclination to be 60\deg).
Thus, even though we have not detected a turnover in the CO rotation
velocity on the northwest side of the galaxy, we believe that the full
width of the CO line is probably a good indicator of the dynamical mass.
The quantitative agreement between the circular velocity derived from CO
and from \oiii\ strongly suggests that the bulk of the
molecular gas 
has relaxed into dynamical equilibrium and is now tracing the same
gravitational potential that the warm gas traces.
Measured \oiii\ velocities extend to larger radii than the CO gas (18\asec\
= 7 kpc) and indicate a dynamical mass of 3.7\e{11} \solmass\ within
7 kpc of the center.

The velocity data also suggest that while CO and \oiii\
may be tracing the same potential, they do not have the same
kinematics.  The ionized gas has a slower rotation speed and a larger velocity
dispersion than the CO. 
The rotation speed of \oiii, corrected for inclination, is 380 \kms\
in the flat part of the rotation curve; its velocity dispersion is
200 \kms\ in the center of the galaxy and 100 \kms\ at 20\asec\ radius
\citep{cretton00}.
In contrast, the CO rotation speed at the edge of the molecular disk 
(also corrected for inclination) is 480 \kms.
The CO velocity dispersion in NGC 2320 cannot be measured directly,
but by analogy with 3C31 and Cen~A we would expect to find a CO
dispersion $\lesssim 20$~\kms\ \citep{okuda05,quillen92}.
Molecular gas could not sustain a velocity dispersion as high as 100--200
\kms\ without being shocked and destroyed.
Of course, it is true that the rotation speed of \oiii\ could be
underestimated by as much as 20\% if the major axis slit were offset from
the kinematic axis by 20\deg, but the large
number of position angles observed by \citet{cretton00} makes this latter
possibility unlikely. 

In contrast to NGC~2320, many spiral galaxies show good agreement between
ionized and molecular gas kinematics.
The Sb galaxy NGC~4527 is a good example \citep{sofue99}.
\citet{lavezzi98a,lavezzi98b} have also shown that in spirals the CO
linewidth is consistent with HI and H$\alpha$ linewidths so that CO
spectra can also be used for Tully-Fisher analyses.
Such agreement between CO and H$\alpha$ kinematics is reasonable if the
ionized gas is in quiescent HII regions which inherited the
global rotation velocity and dispersion of the 
molecular gas from which they formed.
But this is apparently not the case in NGC~2320, where the ionized gas has a
smaller rotation speed and a larger velocity dispersion than the molecular
gas.
We conclude that the \oiii\ in NGC~2320 cannot trace star
formation activity.
Perhaps the warm ionized gas cooled out of a hot ISM phase; an image of \oiii\
emission would be useful for testing this hypothesis.
NGC 2320 has not apparently been searched deeply for X-ray emission, but 
its optical luminosity is well up into the range in which
\citet{osullivan01} find copious amounts of hot gas.

\subsubsection{Angular momentum distribution}\label{2320angmom}

Comparisons of cold gas and stellar kinematics are particularly valuable in
early type galaxies because such comparisons can test the hypothesis that
these galaxies acquired their cold gas from external sources.
In NGC 2320 (Figure \ref{n2320pv}), the maximum rotation velocity of the CO
emission is twice that of the stars.  Interior to 10\asec\ $\sim$  3.8 kpc
$\sim$ $r_e/5$, therefore, the specific angular momentum of the gas is
twice that of the stars.   At face level this result would seem to suggest an external
origin for the gas, but the bulk of the stars in the galaxy are
beyond $r_e/5$ and we should also consider those more distant stars in the
angular momentum comparisons.

We make this analysis of stellar and gaseous angular momenta
through a modified version of a procedure outlined by
\citet{vdbbs01}.   Those authors compared the specific angular momentum
distribution of the baryons in a galaxy to that of a model
dark matter halo as follows.  The specific angular momentum $j \equiv r
v_\phi(r)$ ($v_\phi$
is the mean azimuthal velocity) is a monotonically increasing function of radius,
as long as $v_\phi$ does not drop too quickly. (``Too quickly" would have to
be faster than a Keplerian decline.)  From the mass distribution 
one may also compute $m(r)$, the mass fraction
interior to $r$, which is also a monotonic
function of $r$.  As a result it is
straightforward to use the rotation curve to compute $j(r)$ and to
construct the cumulative specific angular momentum distribution $m(j)$, a
measure of the fractional mass with specific angular momentum less than $j$.
But rather than lumping stars and gas together as \citet{vdbbs01} did, we
wish to consider them separately.

The stellar distribution and kinematics of NGC 2320 are taken from
\citet{cretton00}.  Specifically, we assume an oblate spheroid inclined at
60\deg, the inclination of the gas/dust disk.  The stellar density 
$\rho(R,z)$ is assumed to be the Multi-Gaussian Expansion
model of Cretton et al., including five components with dispersions ranging
from 0.8\asec\ to 56\asec.  We also take the stellar rotation curve, kindly
provided in electronic form by N.\ Cretton, and assume a cylindrical
rotation field (the velocity is independent of the distance $z$ from the
equatorial plane).  
This latter assumption may 
overestimate the stellar angular momenta, but the effect is
probably not great since the assumed Gaussian spheroids have 2:1 axis
ratios.

For the gas, we make three
different models motivated by the facts that (1) the spatial resolution
of the CO data is not great and (2) as yet we have no knowledge of the
distribution of atomic gas.  
The first model neglects HI entirely; it
assumes that the molecular gas is in a disk of constant surface density.
A second model includes HI
by presuming that the gas is in an exponential disk whose total mass is
twice that of the molecular gas and whose scale length is 5\asec.  
The central surface density in this model equals the observed peak \htoo\
column density, 370 \msunsqpc, but the model is too extended in
the sense that the gas surface density is still quite high (50
\msunsqpc) at the observed edge of the CO disk.
A third model has a total mass 1.25 times as large as the H$_2$ mass and a
scale length 3\asec.  This model has a central surface density of 520
\msunsqpc.  It more accurately reproduces the size and mass of the observed 
molecular
disk; its surface density falls to the present sensitivity limit ($3\sigma$
in two consecutive channels = 16 \msunsqpc) at $r = 11''$.
All gas is assumed to follow a rotation curve
which rises linearly to a maximum velocity of 480 \kms\ at a turnover
radius of 5\asec\ and is flat thereafter, based on Figure \ref{n2320pv}.

Figure \ref{angmom} compares the cumulative specific angular momentum
distribution for stars and three gas models of NGC 2320.  
It indicates that the molecular gas has a specific angular
momentum well below that of the bulk of the stars; 30--40\% of the
stellar mass has specific angular momentum greater than the {\it maximum}
value attained by the molecular disk.  This result is independent of 
any assumptions about the distribution of the gas within the disk.
It is also independent of any reasonable assumption about the stellar rotation curve
beyond its last measured point, which happens to be at 
$r \sim 30$\asec\ and $j \sim 1820$~kpc~\kms\ (equal to the maximum angular
momentum attained in the molecular disk).
When reasonable HI disks are added, it is still true that the median
specific angular momentum of the gas (the $j$ value with $m(j) = 0.5$)
is less than that of the stars.
In fact, only the most extreme HI disk assumed above has a tail of high
angular momentum gas which might be inconsistent with the stellar angular
momenta.

This analysis suggests that the angular momentum of the
molecular gas is actually {\it consistent} with an origin in stellar mass
loss.
The mass returning to the ISM would begin, of course, with spatial and
angular momentum distributions matching those of the parent stars.  
The thermal history of the returned mass could be fairly complex, passing
through a hot phase \citep{bm96,bm97}.  As the
gas cooled and settled into a disk it might have contracted, losing a
modest fraction of its angular momentum, into the CO disk which is now
observed.  
Future HI observations will help settle the question of whether any atomic
gas in NGC 2320 is also consistent with an internal origin.

\section{Discussion}

\subsection{Star formation}

The presence of molecular gas naturally suggests the possibility of
star formation.  Most of the star formation activity in
giant ellipticals is supposed to have ended long ago; 
however, recent UV observations suggest that a trickle of star formation
activity may have continued to the present day in some giant ellipticals \citep{yi2005}.
Radio continuum and far-IR fluxes can also be useful as star formation
indicators \citep{condon92,wrobel88}.  In this context, we consider radio continuum
and FIR fluxes of NGC 83, NGC 2320, and NGC 5838 and ask whether evidence
for star formation activity is connected with the presence of molecular
gas in these early type galaxies.

Radio continuum flux densities at 1.4 GHz are available for these three
galaxies from the NRAO--VLA Sky Survey \citep{condon98}.
IRAS 60\micron\ and 100\micron\ flux densities are taken from the NASA
Extragalactic Database (NED), as updated in 1994 by Knapp.
We then calculate a logarithmic FIR/radio flux ratio $q$ as defined by 
\citet{condon92} and others.
The vast majority of gas-rich, star forming spiral galaxies have 
$q \sim 2.34$ with a dispersion of about 0.26 dex \citep{YRC} [YRC].  
Galaxies whose radio emission is dominated by an active galactic nucleus
(AGN) are usually easily identified by their low $q$ values.

Table \ref{q} presents the $q$ values, radio continuum and 
FIR flux densities for NGC 83, NGC 2320, and NGC 5838.
NGC 83 and NGC 5838 lie close to the mean $q$ value of YRC's sample of star forming
galaxies, at 2 and 1 standard deviations in the direction of slight IR
excess.  Since NGC 83 does contain abundant molecular gas, the most 
likely source of the radio and FIR fluxes from
this galaxy is therefore star formation.  The standard interpretation of NGC 5838's $q$ value would also be star formation, though it is curious that
molecular gas has not been reliably detected in NGC 5838.

In contrast, the $q$ value for NGC 2320 is 2.9 standard deviations below
the mean.  (\citet{miller01} find a similar $q$ value to the one derived
here.)
According to the definitions of YRC and \citet{condon02}, NGC 2320 is a ``radio
excess" galaxy meaning that its radio continuum is probably powered by an
AGN.  
That interpretation is consistent with the fact that the radio source
in NGC 2320 is still unresolved by the FIRST survey \citep{white97}, so the
radio source is much smaller in angular size than the molecular gas
distribution.
The kinematic data in Section \ref{2320CO} also imply that most of the
\oiii\ emission from NGC 2320 is not associated with star formation.
Thus, NGC 2320 is rich in molecular gas but there is no conclusive evidence for
ongoing star formation activity.  
NGC 83, also rich in molecular gas, does show evidence for star formation;
and NGC 5838, with no detected molecular gas, shows radio continuum and FIR
fluxes which would normally be interpreted as star formation.
Evidently the connection between molecular gas, star formation, and
radio/FIR emission is not as straightforward in early type galaxies as it
is in late type galaxies.

\citet{okuda05} and \citet{koda05} have shown that the molecular surface densities in early type
galaxies can be quite high --- as high as in starburst galaxies ---
without inducing detectable star formation.
The molecular disks apparently are stabilized by their host galaxies' large
dynamical masses, as one can see from an analysis of Toomre's $Q$ parameter
\citep{kennicutt1989}.
The $Q$ parameter is intended to estimate the degree of gravitational
stability of a gas disk in a galaxy, as it is a ratio of the
properties which inhibit the growth of gravitational instabilities (local
velocity dispersion and epicycle frequency) to the properties which enhance
such growth (the gas surface density).
\citet{koda05} have rewritten Toomre's $Q$ parameter
as $Q \sim \alpha\frac{(M_\sigma M_{dyn})^{1/2}}{M_{gas}},$
where $M_\sigma \equiv R\sigma^2/G,$ $\sigma$ is the local gas velocity
dispersion, and $\alpha$ encapsulates the
information about the local shape of the rotation curve.
When the $Q$ parameter is written in this way it is easy to see that
for two galaxies with similar molecular disks (similar $R$, $\sigma$, and
$M_{gas}$ or $\Sigma_{gas}$), the one with smaller $M_{gas}/M_{dyn}$ has a
larger Toomre $Q$ and should have less star formation activity.

In this context, the molecular disks of NGC~83 and 
NGC~2320 make interesting comparisons with similar disks in 3C~31
\citep{okuda05} and early-type spirals \citep{koda05}.
As in 3C~31, the molecular surface densities in NGC~83 and NGC~2320 are
respectably high; the peak values are 140 \solmass~pc$^{-2}$ and 370
\solmass~pc$^{-2}$, not including helium.
The present data do not permit a meaningful estimate of Toomre's $Q$ for the
molecular disks in NGC~83 and NGC~2320, but it is clear that these galaxies 
have low ratios of gas to dynamical mass.
Specifically, $M_{gas}/M_{dyn} \sim 0.03$ in NGC~83 and NGC~2320, similar to
the value of 0.02 found in 3C~31 \citep{okuda05}.
Late-type spirals typically have $M_{gas}/M_{dyn} \sim 0.1$ to 1.0 \citep{koda05}.
Thus the rather low $M_{gas}/M_{dyn}$ in NGC~83 and NGC~2320 could provide
a plausible explanation for a lack of star formation in these disks.

\subsection{Disk growth}

Even if star formation does transform the molecular disks into stellar
disks, the future stellar disks will not be large or dramatic.
The radial extent of the CO emission in NGC 83 and NGC 2320 only
corresponds to $\sim$  0.2~$r_e$ (where $r_e$ is the effective radius or
the half-light radius of the de Vaucoleurs $r^{1/4}$ profile).
The CO disks in NGC 83 and NGC 2320 are thus smaller, relatively speaking,
than the ones in the early type galaxies
NGC~807, UGC~1503, NGC~3656, NGC~4476, and NGC~5666 \citep{young02}.
Those latter five CO disks extended to 0.5--1~$r_e$.

The CO disks in NGC 83 and NGC 2320 are also small when measured as mass
fractions of their host galaxies.
\citet{cretton00} quote $M_V = -22.5$ for NGC 2320 with a dynamical
$V$-band mass-to-light ratio of 16,  which means that the estimated
molecular gas mass of 4.3\e{9} \solmass\ is only 0.3\% of the stellar mass.
Similarly, if NGC~83 has $M_V = -22.1$ and $(M/L)_V \sim 10$ then its molecular
mass (2.0\e{9} \solmass) is only 0.3\% of its stellar mass.
Star formation might transform CO-rich elliptical galaxies into
``two-component disky ellipticals"
of the type discussed by \citet{khochfar04}, but any
future stellar disks in NGC~83 and NGC~2320 
would be significantly smaller in
radial extent, less luminous, and more difficult to detect than the
embedded stellar disks studied by \citet{scorza95} and \citet{scorza98},
for example.

On the other hand,
the molecular disk in NGC 2320 is {\it already} associated with an embedded 
stellar disk.  
Stellar velocity profiles show that the 
the Gauss-hermite $h_3$ parameter is anticorrelated with the mean rotation 
velocity \citep{cretton00}; this is the 
classic signature of a dynamically cold disk
embedded in a dynamically hot galaxy \citep{vdmf93}.
This kinematic disk extends to at least 20\asec\ and thus is coincident with
the bright annulus at 15\asec\ (Section \ref{2320optical}).
The CO disk in NGC 83 is not, apparently, associated with a stellar disk:
in Section \ref{83optical} we found no morphological
evidence for an embedded stellar disk beyond the obvious dust disk, though
it would be useful to check stellar kinematics of this galaxy.
Perhaps, then, the molecular gas in NGC 2320 is a small remnant of a
once-larger gas disk which formed the dynamically cold stellar disk.
Further insight into this question should come with more detailed mapping of
the stellar populations and with an understanding of the origin of the
molecular gas.

\subsection{The origin of the cold gas}

Since the gas and dust in NGC 83 appear so well relaxed, we conclude that
{\it if} they were acquired from an external source it must have been
several orbital timescales ago.
The orbital timescale at the edge of the
CO/dust disk is 3\e{7}--4\e{7} yr (Section \ref{83CO}).
The general symmetric appearance of the optical images also suggests that
it has been well over a Gyr since NGC 83 experienced any merger, if there was one.
Stellar kinematic maps of NGC 83 will be necessary before we can say more
about whether the angular momentum distribution of the stars is really
inconsistent with that of the gas.

In contrast with NGC 83, the 
the asymmetries previously noted in the CO distribution of NGC 2320 suggest
that not all of the molecular gas in this galaxy is fully relaxed.
The general appearance of the gas and dust 10\asec\
southeast of the nucleus is that of gas which has not yet settled into
a relaxed disk (Figure \ref{n2320dust} and \ref{n2320channelsb}).
On the other hand, in the position-velocity diagram (Figure \ref{n2320pv})
this ``extra" gas at 6240 \kms\ appears to be a natural turnover in the CO
rotation curve.  Thus all of the detected CO {\it could} be
relaxed into dynamical equilibrium, but then it is curious that the radial
extent and the luminosity of the molecular gas are twice as large on the
southeast side of the galaxy as on the northwest side.
Regardless of whether the CO emission at 6240 \kms\ is interpreted as
gas in or out of dynamical equilibrium, we are led to the conclusion that
some unusual event must have happened in the core of NGC 2320 a short time
ago.  The orbital timescale at 10\asec\ radius and 480 \kms\ is only 5\e{7}
yr; strong side-to-side asymmetries in the gas distribution should have
been smeared out over a few orbital timescales.

The comparisons of the specific angular momentum distributions of gas and
stars in NGC 2320 (Section \ref{2320angmom}) showed that the presently
observed CO disk might conceivably have its origin from internal stellar
mass loss.  Naturally, such comparisons cannot rule out an external origin,
since the angular momentum which is left after a minor or a major merger
depends a great deal on the (unknown) geometry of the interaction.
Additional studies of larger
numbers of CO-rich early type galaxies will be necessary before we can hope
to piece together a coherent picture of these fascinating systems.

\section{Summary}

We present a study of the cold ISM in 
NGC 83, NGC 2320, and NGC 5838, three early type
galaxies selected from single dish CO surveys.
Broadband optical imaging of NGC 83 and NGC 2320 gives their optical and
dust morphologies.
New observations with the BIMA millimeter array give a sensitive CO 
limit for NGC 5838 as well as the distribution and kinematics of molecular
gas in NGC 83 and NGC 2320.

The optical images of NGC 83 and NGC 2320 show classic $r^{1/4}$ surface
brightness profiles, lending support to the contention that these are true
giant elliptical galaxies rather than misclassifed spirals.
In both of these galaxies the CO emission is very closely coincident with
internal dust disks (the same is not true for NGC 5838, which also has a dust
disk but no detected CO).
The kinematic axes of the molecular disks are aligned within a few degrees of the
large scale optical major axes, as is also true for most of the
early-type galaxies studied by \citet{young02}.  The alignment suggests 
that these galaxies are probably oblate spheroids.

We also calculate dynamical masses within the outer edge of the CO disks
and find that they are consistent with little, if any, dark matter in the
interiors of these galaxies; this result is consistent with (but
independent of) previous work
on stellar kinematics in early-type galaxies.

The CO kinematics also provide clues to the origin of the gas.
The kinematics of the CO in NGC 83 suggest that the gas/dust disk is
well settled into dynamical equilibrium; if this gas was acquired from an external
source, it must have been much longer than about 4\e{7} yr ago.
Most of the CO emission and dust in NGC 2320 also appears to be settled into a
relaxed disk.
CO kinematics in NGC 2320 are consistent with the circular velocity curve
inferred by \citet{cretton00} from \oiii\ emission. 
However, 25\% of the CO is in an asymmetric structure beyond one
edge of the disk.  
This asymmetric structure should be sheared by differential rotation on an
orbital timescale of 5\e{7} yr and the short orbital timescale suggests that
the gas was acquired or disturbed recently.

We compare the cumulative specific angular momentum distribution of the CO
in NGC 2320 to that of the stars.  At a given radius (within the CO/dust
disk), the rotation velocity of the gas is at least twice the mean velocity
of the stars.  But since the stars extend so much farther than the CO does,
under plausible approximations the specific angular momentum distribution of
the cold gas is not inconsistent with that of the stars.  From an angular
momentum 
point of view, the molecular gas in NGC~2320 could conceivably have
originated in internal stellar mass loss, contraction, and spin-up.

Even though NGC 83 and NGC 2320 are unusually rich (for early-type
galaxies) in molecular gas, evidence for star formation activity and
stellar disk growth is mixed.  For example,
the radio and FIR fluxes in NGC 83 suggest possible star formation activity
but no unusual structure is noted in the optical morphology.
There is a dynamically cold stellar disk or ring in NGC 2320, with the same
orientation and a bit larger radial extent than the molecular disk;
however, there seems to be little present-day star formation in NGC 2320.
The galaxy's radio and FIR fluxes suggest that its radio emission is
dominated by an AGN rather than by star formation.
Furthermore, the ionized gas in NGC 2320 has
significantly higher velocity dispersion and lower rotation velocity than
the CO does.  That \oiii\ emission does NOT trace star formation
activity in HII regions formed out of the molecular gas; it must have some
other energizing source, and perhaps it is cooling out of a hot gas phase.
In any case, if the molecular disks in NGC 2320 and NGC 83 do form stellar
disks they will be smaller both in relative mass and in radial extent than
the ones which are commonly found in disky ellipticals.

\acknowledgments

Thanks to Tom Statler and Jacqueline van Gorkom for helpful
comments and to Liese van Zee for the optical images.
Thanks also to the Berkeley-Illinois-Maryland Association 
(operated with support from the National
Science Foundation) for generous investments of telescope time.
This work was partially supported by NSF AST-0507423 and it 
has made use of the NASA/IPAC Extragalactic Database (NED) which is operated by the Jet Propulsion Laboratory, California Institute of Technology, under contract with the National Aeronautics and Space Administration. 

Facilities: \facility{BIMA}, \facility{WIYN}.

\clearpage

\begin{figure}
\includegraphics[scale=1.3]{f14_color.eps}
\figurenum{14}
\caption{NGC 2320 major axis position-velocity diagram.  
Greyscale and contours show the CO intensity at $-20$, 20, 40, 60, and 80
percent of the peak (64 \mjb).
The solid line (red crosses in the electronic edition) indicates velocities measured from the stellar absorption
line data of \citet{cretton00}.
The dotted line (blue crosses) indicates [OIII] velocities from the same source.
All of these are uncorrected for inclination, i.e. they are $V \sin i$.
The long-dash line (green crosses) indicates the circular velocity curve derived by Cretton
\etal\ by correcting the [OIII] velocities for inclination and asymmetric
drift.  Here we have multiplied the circular velocities by $(\sin i)$ in
order to compare with the observed CO kinematics. In the
limit that the velocity dispersion of the molecular gas is very small and the
CO has relaxed into dynamical equilibrium, 
the CO emission is expected to follow the circular velocity curve.
\label{n2320pv}
}
\end{figure}

\clearpage
\begin{figure}
\includegraphics[scale=0.7]{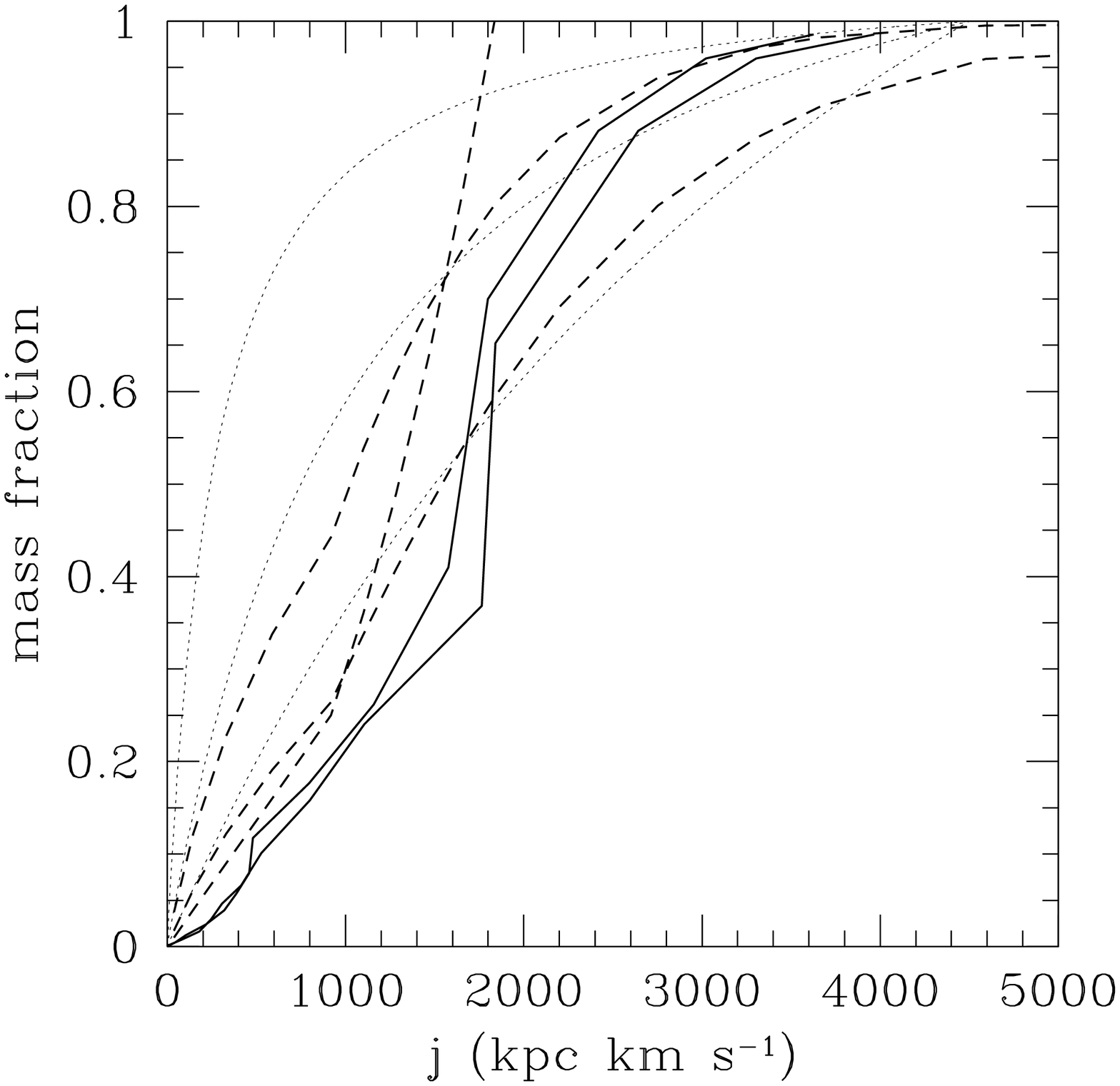}
\figurenum{15}
\caption{Cumulative specific angular momenta of gas and stars in NGC~2320.
The vertical axis plots the mass fraction $m(j)$ with specific angular
momentum less than $j$.  Two heavy solid lines show $m(j)$ calculated as
described in the text for the stellar rotation (both sides of the major
axis are plotted).  The difference between the two solid lines is an
indicator of the uncertainty introduced by errors in the stellar
velocities.  Three heavy dashed lines show $m(j)$ for the cold gas in the
galaxy; the model which includes only \htoo\ reaches $m(j)=1$ at the edge
of the observed CO disk where $j = 1800$ kpc~\kms.  The other two dashed
lines include hypothetical extended atomic hydrogen disks.
For comparison,
three light dotted lines show the angular momentum distributions of
$\Lambda$CDM dark matter halos as in \citet{vdbbs01}.  A maximum specific
angular momentum $j_{max} = 4500$ kpc~\kms\ is assumed for these dark
matter halos.
\label{angmom}
}
\end{figure}
\clearpage
\begin{deluxetable}{lccrc}
\tablewidth{0pt}
\tablecaption{Sample Galaxies -- Optical Properties
\label{sampletable}}
\tablehead{
\colhead{} & \colhead{NGC 83} & \colhead{NGC 2320} & \colhead{NGC 5838} \\
}
\startdata
RA (J2000.0) & 00 21 22.4  & 07 05 42.0  & 15 05 26.2 \\
Dec &  +22 26 01 & +50 34 52 & +02 05 58  \\
Velocity (\kms) &  6359 (27) & 5944 (15) & 1359 (10) \\
Distance (Mpc) & 85  & 79 & 18.7  \\
$M_B$ &   $-21.1$   & $-21.5$   & $-19.6$  \\
$(B-V)_e$ & 1.12 & 1.05 & 1.02 \\
$\sigma_0$ (\kms) & 250 & 350 &  266 \\
$r_e$ ($''$) &  27 &  43 &  \nodata \\
$r_e$ (kpc) &  11  & 16 & \nodata \\
\enddata
\tablecomments{Velocities and blue magnitudes are taken from the NASA Extragalactic Database
(NED). 
The distance estimates for NGC 83 and NGC 2320 are taken from WCH95, who used 
$H_0 = 75$ \kms~Mpc$^{-1}$ and a Virgocentric infall
model.
The distance for NGC 5838 is from \citet{dezeeuw02}.
Colors and velocity dispersions are from the hyperLEDA database; effective
radii are from \citet{burstein87}.
}
\end{deluxetable}

\begin{deluxetable}{lccccc}
\tablewidth{0pt}
\tablecaption{CO Observation Parameters
\label{obstable}}
\tablehead{
\colhead{Galaxy} & \colhead{Flux cal} & \colhead{Phase cal} &
\colhead{Velocity Range} & \colhead{FOV} 
& \colhead{3mm cont.} \\
\colhead{} & \colhead{} & \colhead{} & \colhead{\kms} & \colhead{kpc} & \colhead{mJy}
}
\startdata
NGC 83  & Uranus, 3C454.3 & 3C454.3 & 5673--6986 & 41 & $<$ 5.1 \\
NGC 2320 & Mars & 0646+448, 0533+483, & 5160--6490 & 38 & $<$ 13.5 \\
         &      &     0753+538        &            &    &          \\
NGC 5838 &  Mars & 1550+054 & 669--1960 & 9 & $<$ 8.1 \\
\enddata
\tablecomments{Field of view (FOV) is the FWHM of the primary beam (100\asec) at
the distances in Table \ref{sampletable}.  
}
\end{deluxetable}

\begin{deluxetable}{lccccc}
\tablewidth{0pt}
\tablecaption{CO Image Properties
\label{imgtable}}
\tablehead{
\colhead{Galaxy} &
\colhead{Beam} & \colhead{Beam} & 
\colhead{Channel} & \colhead{noise} & \colhead{N(H$_2$) limit} \\
\colhead{} & \colhead{\asec} & \colhead{kpc} &
\colhead{\kms} & \colhead{\mjb} & \colhead{\tenup{20} cm$^{-2}$}
}
\startdata
NGC 83  & 6.6$\times$6.0 & 2.7$\times$2.5 & 20.9 & 7.7 & 3.3 \\
         & 6.6$\times$6.0 & 2.7$\times$2.5 & 41.7 & 5.8 & 5.1 \\
         & 10.1$\times$9.7 & 4.2$\times$4.0 & 41.7 & 7.4 & 2.6 \\
NGC 2320 & 6.4$\times$5.6 & 2.4$\times$2.2 & 20.8 & 9.9 & 4.7 \\
         & 6.4$\times$5.6 & 2.4$\times$2.2 & 41.6 & 7.5 & 7.1 \\
         & 9.8$\times$9.4 & 3.7$\times$3.6 & 41.6 & 9.8 & 3.7 \\
NGC 5838 & 8.1$\times$6.0 & 0.83$\times$0.55 & 20.2 & 19 & 6.5 \\
         &  11.3$\times$10.0 & 1.0$\times$0.9 & 40.3 & 17 & 4.9 \\
\enddata
\tablecomments{The N(H$_2$) limit is a sensitivity estimate corresponding
to a 3$\sigma$ signal in one channel. }
\end{deluxetable}

\begin{deluxetable}{llll}
\tablewidth{0pt}
\tablecaption{FIR and Radio Continuum Emission
\label{q}}
\tablehead{
\colhead{} & \colhead{NGC 83} & \colhead{NGC 2320} &
\colhead{NGC 5838} 
}
\startdata
$S_{100 \micron}$, Jy & 2.15\error 0.18 & 1.60\error 0.10 & 1.67\error 0.09 \\
$S_{60 \micron}$, Jy & 0.34\error 0.09 & 0.26\error 0.02 & 0.73\error 0.04 \\
$S_{1.4 GHz}$, mJy & $<$ 1.5 & 19.3\error 0.7 & 2.6\error 0.4 \\
$q$                & $>$ 2.83 & 1.60\error 0.03 & 2.66\error 0.07 \\
\enddata
\end{deluxetable}

\end{document}